\documentclass[a4paper, aps, prd, showpacs, superscriptaddress, groupedaddress, preprintnumbers, nofootinbib, twocolumn]{revtex4-1} 
\usepackage[pdftex]{graphicx, color}   
\usepackage{amsmath, amssymb, amscd, latexsym, bm, braket}   
\usepackage{url}
\DeclareMathAlphabet{\mathpzc}{OT1}{pzc}{m}{it}
\newcommand{\Slash}[1]{{\ooalign{\hfil/\hfil\crcr$#1$}}}

\usepackage{color}

\begin{document}
\title{Franz-Keldysh effect in strong-field QED}
\author{Hidetoshi Taya}
\email{{\tt h_taya@fudan.edu.cn}}
\affiliation{Department of Physics and Center for Field Theory and Particle Physics, Fudan University, Shanghai, 200433, China }

\date{\today}

\begin{abstract}
We studied a QED analog of the Franz-Keldysh effect, and the interplay between the non-perturbative (the Schwinger mechanism) and the perturbative particle production mechanism from the vacuum in the presence of a strong slow field superimposed by a weak field.  We found that the Franz-Keldysh effect significantly affects the particle production: (i) the perturbative particle production occurs even below the threshold energy; (ii) the perturbative production becomes the most efficient just above the threshold energy; and (iii) an oscillating behavior appears in the production number above the threshold energy.  These non-trivial changes are suppressed only weakly by powers of the critical field strength of QED.  A relation to the dynamically assisted Schwinger mechanism and implications to experiments are also discussed.  
\end{abstract}


\maketitle

\section{Introduction}

It was Dirac who first discovered a relativistic wave equation for electron, which is known as the Dirac equation today \cite{dir28}.  The Dirac equation admits infinitely negative energy states.  This looks problematic because any state may fall into lower and lower energy states by emitting photons and thus there seem no stable states.  This problem was resolved by Dirac himself by re-interpreting that negative energy states are all occupied in our physical vacuum (Dirac sea picture) \cite{dir30}.  Dirac's interpretation suggests that our vacuum is not vacant space, but can be regarded as something like a semi-conductor with gap energy characterized by the electron mass scale.  This implies that our vacuum exhibits non-trivial responses when exposed to external fields whose characteristic physical scale is larger than the gap energy, as semi-conductors do.

One of the most interesting responses is particle production from the vacuum in the presence of external electric fields.  Roughly, there are two kinds of production mechanism, whose interplay is controlled by strength and frequency of the external field (or the Keldysh parameter) \cite{kel65, bre70, pop72, tay14}.  

Namely, the first mechanism is the perturbative production mechanism, which occurs when the external field is weak but has high-frequency (i.e., energetic).  This is an analog of the photo-absorption effect in semi-conductors.  In this mechanism, the external field perturbatively kicks an electron filling the Dirac sea, and supplies energy.  If the supplied energy (i.e., the frequency of the external field) is larger than the gap energy, the electron is kicked out to the positive energy band leaving a hole in the original negative energy state.  Thus, a pair of an electron and a positron is produced.  This mechanism is suppressed only weakly by powers of the coupling constant $e$.  Thus, it is not difficult to study the perturbative production mechanism with actual experiments (e.g. SLAC E144 experiment \cite{bam99}).

The other mechanism is the non-perturbative production mechanism, which is the so-called Schwinger mechanism \cite{sau31, hei36, sch51}.  This can be understood as an analog of the electrical breakdown of semi-conductors (or the Landau-Zener transition \cite{lan32, zen32, stu32, maj32}) \cite{lin18}.  This mechanism occurs when the external field is strong but has low-frequency.  In the presence of strong electric field, the energy bands are tilted non-perturbatively, and a level crossing occurs.  An electron filling the Dirac sea is now able to tunnel into the positive energy band, which results in spontaneous pair production of an electron and a positron.  If the external field is slow enough, one may approximate the external field as a constant electric field.  For this case, one can analytically derive a formula for the number of produced electrons (the Schwinger formula \cite{sch51}) as 
\begin{align}
	n_{{\bm p},s}^{\rm (Sch)} =  \frac{V}{(2\pi)^3} \exp\left[  -\pi \frac{m_{\rm e}^2 + {\bm p}_{\perp}^2}{eE}  \right], \label{eq1-1}
\end{align}
where ${\bm p}_{\perp}$ is transverse momentum with respect to the direction of the electric field.  As apparent from Eq.~(\ref{eq1-1}), the Schwinger mechanism depends on the coupling constant $e$ inversely in the exponential.  This clearly shows the non-perturbative nature of the Schwinger mechanism.

Experimental verification of the Schwinger mechanism is very important and interesting because it opens up a novel way to unveil non-perturbative aspects of quantum electrodynamics (QED), which are one of the most unexplored areas of modern physics.  Nevertheless, this has not been done yet.  This is because the Schwinger mechanism is strongly suppressed by the exponential factor in Eq.~(\ref{eq1-1}).  Thus, it requires extremely strong electric field $eE_{\rm cr} \equiv m_{\rm e}^2 \sim \sqrt{10^{28}\;{\rm W/cm^2} }$ to be manifest.  Unfortunately, such a strong electric field is not available within our current experimental technologies.  For example, HERCULES laser holds the present world record for highest-intensity focused laser, whose strength is $eE \sim \sqrt{10^{22}\;{\rm W/cm^2}}$ \cite{yan08}.  Upcoming intense laser facilities such as ELI \cite{eli} and HiPER \cite{hiper} may reach $eE \sim \sqrt{ 10^{24}\;{\rm W/cm^2}}$, which is still weaker than the critical field strength $E_{\rm cr}$ by several orders of magnitude.

In recent years, there has been an increasing interest in a ``cooperative'' particle production mechanism between the perturbative and the non-perturbative mechanism.  Namely, the dynamically assisted Schwinger mechanism \cite{sch08, dun09, piz09, mon10a, mon10b} is attracting much attention.  The dynamically assisted Schwinger mechanism claims that the non-perturbative particle production (the Schwinger mechanism) by a strong slow electric field should be dramatically enhanced if one superimposes a weak fast (i.e., perturbative) electric field at the same time onto the vacuum.  Note that it is usually assumed that the perturbative electric field is fast but with a frequency that is still below the electron mass scale.  An intuitive explanation of this mechanism is the following: Firstly, a perturbative interaction kicks up an electron in the Dirac sea into the gap.  Then, the electron inside of the gap is able to tunnel into the positive energy band easier because the tunneling length is reduced compared to the original length from the negative energy band.  One of the striking results of the dynamically assisted Schwinger mechanism is that the critical field strength $E_{\rm cr}$ is reduced by several orders of magnitude by the perturbative kick.  It is, thus, expected that this mechanism might be detectable even within the current experimental technologies.  Although experimental verification of the dynamically assisted Schwinger mechanism is not equivalent to direct verification of the original Schwinger mechanism, it is still very interesting and important because it clearly involves non-perturbative aspects of QED.

Is there any other cooperative particle production mechanism?  If there is, it should serve as another powerful tool to investigate non-perturbative aspects of QED just like the dynamically assisted Schwinger mechanism does.  In the area of semi-conductor physics, there is.  This is the Franz-Keldysh effect \cite{fra58, kel58, tah63, cal63}.  The Franz-Keldysh effect states that optical properties of bulk semi-conductors are significantly modified in the presence of a strong slow electric field.  Namely, photo-absorption rate (i.e., the perturbative particle production rate) under a strong slow electric field becomes finite even below the gap energy, and exhibits an oscillating behavior above the gap energy.  These non-trivial changes are caused by non-perturbative interactions between valence-band electrons and the strong electric field (which will be explained later in more detail in the language of QED).  One of the most important features of the Franz-Keldysh effect is that its suppression is not so strong although it involves non-perturbative physics, which is usually strongly suppressed and is hard to study with experiments.  Thanks to this advantage, the Franz-Keldysh effect in semi-conductors has been tested extensively by numerous experiments since 1960's \cite{boe59, wil60, wil62, vav60, mos61}, and has many applications ranging from physics to industry (e.g. electro-absorption modulator, photo-detector, optical switching, etc).

Since semi-conductors are quite analogous to QED according to Dirac's interpretation, it may be natural to ask if there is an analog of the Franz-Keldysh effect in QED.  To the best of our knowledge, there exists no clear answer to this question.  The purpose of this paper is to answer this question.  Namely, we consider a situation such that a weak field with arbitrary frequency is applied onto a strong slow field, and discuss how the perturbative particle production by the weak field is modified by the strong field.  We shall show that a QED analog of the Franz-Keldysh effect actually takes place, and non-trivial changes in the perturbative production number appear such as excess below the gap energy, and an oscillating behavior above the gap energy.  These changes are found to be suppressed only weakly by powers of the critical field strength.  We shall also clarify a relation between the Franz-Keldysh effect and the dynamically assisted Schwinger mechanism, whose physical set-ups are similar to each other.  In addition to these, we shall study the interplay between the perturbative and the non-perturbative particle production mechanism to clarify the cooperative nature of the Franz-Keldysh effect.

Note that the weak field is not necessarily on-shell here.  Thus, the weak field alone is able to produce particles from the vacuum by the perturbative mechanism.  This situation is in contrast to stimulated pair production processes by an on-shell photon in the presence of strong fields (e.g. non-linear Breit-Wheeler process \cite{ritus}).  An on-shell photon alone is not able to produce particles from the vacuum because of the energy-momentum conservation.  Hence, such stimulated pair production processes cannot be regarded as a cooperative particle production mechanism between the perturbative and the non-peturbative production mechanism, in which we are interested.

This paper is organized as follows: In Sec.~\ref{sec2}, we explain a theoretical framework of this work.  To be concrete, we consider a general situation, where a weak field with arbitrary frequency is superimposed onto a strong slow field.  We derive a general expression for the number of produced particles from the vacuum under those fields by developing a perturbation theory, in which interactions due to the weak field are treated perturbatively but those due to the strong field are treated non-perturbatively.  Note that a similar perturbative approach was recently developed in Ref.~\cite{tor17} in the context of the dynamically assisted Schwinger mechanism\footnote{There are several differences between Ref.~\cite{tor17} and our perturbation theory.  Ref.~\cite{tor17} computes a matrix element for a single pair production process from the vacuum on the basis of the standard $S$-matrix formalism.  Ref.~\cite{tor17} also uses the WKB approximation and a weak field approximation in evaluating the first order perturbative correction to the matrix element, in which the Bogoliubov transformation between in- and out-state annihilation/creation operators is approximated to be an identity matrix (i.e., $\alpha_{\bm p} \to 1, \beta_{\bm p} \to 0$).  On the other hand, we shall directly compute the expectation value of the number operator by using the retarded Green function method without any approximations.  Note that the expectation value of the number operator and the (square of) the matrix element for a single production process are, strictly speaking, not the same quantity because the former gives the inclusive production number, which includes not only a single production process but also multiple production processes \cite{bal01}.  }.  In Sec.~\ref{sec3}, we consider a specific field configuration to quantitatively discuss the particle production mechanism in the presence of both strong and weak fields.  Namely, we consider a constant homogeneous strong electric field and a monochromatic weak electric field.  Based on our perturbation theory, we derive an analytical formula (without any approximations such as the WKB approximation) for the number of produced particles for this particular field configuration.  With this formula, we explicitly demonstrate how a QED analog of the Franz-Keldysh effect and the interplay between the non-perturbative and the perturbative particle production occur.  We also clarify a relation between the Franz-Keldysh effect and the dynamically assisted Schwinger mechanism.  Section~\ref{sec4} is devoted to summary and discussion.

\section{Formalism} \label{sec2}

In this section, we shall derive a formula for the number of particles produced from the vacuum in the presence of a strong slow field and a small perturbation on top of it.  We first use the retarded Green function technique to solve the Dirac equation perturbatively with respect to the perturbation, while we treat interactions due to the strong field non-perturbatively (Sec.~\ref{sec2-a}).  Then, we canonically quantize the field operator (Sec.~\ref{sec2-b}), and directly compute the in-vacuum expectation value of the number operator (Sec.~\ref{sec2-c}).

Note that we use the mostly minus metric $g^{\mu\nu} = {\rm diag}(+1,-1,-1,-1)$.  Also, we work in the Heisenberg picture throughout this paper.

\subsection{Perturbative solution of the Dirac equation} \label{sec2-a}

We consider a situation such that an external gauge field $A_{\mu}$ can be separated into two parts, i.e., a strong and slow field $\bar{A}_{\mu}$ and a weak field ${\mathcal A}_{\mu}$, which is applied as a perturbation on top of the strong field $\bar{A}_{\mu}$, as 
\begin{align}
	A_{\mu} = \bar{A}_{\mu} + {\mathcal A}_{\mu}.  
\end{align}
We assume that the weak field ${\mathcal A}_{\mu}$ vanishes at the infinite past and future (adiabatic hypothesis) as
\begin{align}
	{\mathcal A}_{\mu} \xrightarrow{x^0 \to \pm \infty}{} 0.    \label{eqq3}
\end{align}
For simplicity, we adopt the temporal gauge fixing condition, i.e., 
\begin{align}
	\bar{A}_{\mu} = (0, -\bar{\bm A}),\  {\mathcal A}_{\mu} = (0, -{\bm {\mathcal A}}), 
\end{align}
where we introduced the three-vector potential $\bar{\bm A}, {\bm {\mathcal A}}$ as the spatial component of the corresponding gauge field.

Under the external field, the Dirac equation for a fermion field operator $\hat{\psi}$ reads
\begin{align}
	0 &= \left[ i\Slash{\partial} - e\Slash{A} -m \right] \hat{\psi} \nonumber\\
	  &= \left[ i\Slash{\partial} - e\bar{\Slash{A}} -m \right] \hat{\psi} - e\Slash{\mathcal A}\hat{\psi} , \label{eq1-2}
\end{align}
where $e>0$ is the coupling constant and $m$ is mass.  Now, we shall solve the Dirac equation (\ref{eq1-2}) perturbatively with respect to ${\mathcal A}_{\mu}$, while interactions between $\bar{A}_{\mu}$ and $\hat{\psi}$ are treated non-perturbatively.  To this end, we introduce a retarded Green function $S_{\rm R}$ such that 
\begin{align}
	&\left[ i\Slash{\partial}(x) - e\bar{\Slash{A}}(x) -m \right] S_{\rm R}(x,y) = \delta^4(x-y),\nonumber\\
	&S_{\rm R}(x,y) = 0\ \ {\rm for}\ \ x^0-y^0<0 . 
\end{align}
Notice that $S_{\rm R}$ is fully dressed by the strong field $\bar{A}_{\mu}$ (Furry picture \cite{fur51}).  With the Green function $S_{\rm R}$, one can write down a formal solution of the Dirac equation (\ref{eq1-2}) as
\begin{align}
	\hat{\psi}(x) 
	 &= \sqrt{Z} \hat{\psi}^{\rm in}(x) + e \int d^4 y S_{\rm R}(x,y) \Slash{\mathcal A}(y) \hat{\psi}(y) \nonumber\\
	 &=  \sqrt{Z} \left[ \hat{\psi}^{\rm in}(x) + e \int d^4 y S_{\rm R}(x,y) \Slash{\mathcal A}(y) \hat{\psi}^{\rm in}(y) + {\mathcal O}(e^2) \right] , \label{eq1-4}
\end{align}
where we used Eq.~(\ref{eqq3}) and imposed a boundary condition for the field operator $\hat{\psi}$ (Lehmann-Symanzik-Zimmermann (LSZ) asymptotic condition \cite{leh55}) as\footnote{Strictly speaking, the equality in Eq.~(\ref{eq1-6}) should be interpreted in a weak sense, i.e., the equality holds only after (products of) operators are sandwiched by states.  This difference is not important in the following discussion since we are basically interested in the expectation value of number operator.  }
\begin{align}
	0 = \lim_{x^0 \to -\infty}\left[ \hat{\psi} - \sqrt{Z} \hat{\psi}^{\rm in}  \right]. \label{eq1-6}
\end{align}
Here, $Z = 1 +{\mathcal O}(e)$ is a field renormalization constant and $\hat{\psi}^{\rm in}$ is a solution of the Dirac equation without the weak field ${\mathcal A}_{\mu}$ such that 
\begin{align}
	0 = \left[ i\Slash{\partial} - e\bar{\Slash{A}} -m \right]  \hat{\psi}^{\rm in}.  \label{eq1-7}
\end{align}

\subsection{Annihilation/creation operators} \label{sec2-b}

One can define an annihilation/creation operator at in-state $x^0 \to -\infty$ by canonically quantizing the asymptotic field operator $\hat{\psi}^{\rm in}$.  To be more concrete, we first expand the asymptotic field operator $\hat{\psi}^{\rm in}$ in terms of a mode function ${}_{\pm} \psi^{\rm in}_{{\bm p},s}$ as
\begin{align}
	\hat{\psi}^{\rm in}(x) = \sum_s \int d^3{\bm p} \left[ {}_+ \psi^{\rm in}_{{\bm p},s}(x) \hat{a}^{\rm in}_{{\bm p},s} + {}_- \psi^{\rm in}_{{\bm p},s}(x) \hat{b}^{{\rm in}\dagger}_{-{\bm p},s}   \right]
\end{align}
with ${\bm p}$ and $s$ being a label of canonical momentum and spin, respectively.  Here, we normalize the mode function ${}_{\pm} \psi^{\rm in}_{{\bm p},s}$ by
\begin{align}
	\int d^3{\bm x} {}_{\pm} \psi^{{\rm in}\dagger}_{{\bm p},s} {}_{\pm} \psi^{\rm in}_{{\bm p}',s'} &= \delta^3({\bm p}-{\bm p}')\delta_{ss'},\nonumber\\
	\int d^3{\bm x} {}_{\pm} \psi^{{\rm in}\dagger}_{{\bm p},s} {}_{\mp} \psi^{\rm in}_{{\bm p}',s'} &= 0, \label{eq1-8}
\end{align}
and identify the positive/negative frequency mode if it approaches a plane wave with positive/negative frequency at $x^0 \to -\infty$ as
\begin{align}
	\lim_{x^0 \to -\infty} {}_{\pm} \psi^{\rm in}_{{\bm p},s} \propto {\rm e}^{\mp i\omega_{ {\bm P}^{\rm in}} x^0} {\rm e}^{+i {\bm p}\cdot{\bm x}}, \label{eq1-9}
\end{align}
where
\begin{align}
	\omega_{\bm p} \equiv \sqrt{m^2 + {\bm p}^2}
\end{align}
is on-shell energy, and ${\bm P}^{\rm in} \equiv {\bm p}-e\bar{\bm A}(x^0 = -\infty)$ is kinetic momentum at $x^0 \to -\infty$.  Nextly, we impose the canonical commutation relation onto $\hat{\psi}^{\rm in}$.  This is equivalent to quantize $\hat{a}^{\rm in}, \hat{b}^{{\rm in}\dagger}$ as
\begin{align}
	&\delta^3({\bm p}-{\bm p}')\delta_{ss'} = \{ \hat{a}_{{\bm p},s}^{{\rm in}\dagger} ,  \hat{a}_{{\bm p}',s'}^{{\rm in}}  \} = \{ \hat{b}_{{\bm p},s}^{{\rm in}\dagger} ,  \hat{b}_{{\bm p}',s'}^{{\rm in}}  \},  \nonumber\\
	&({\rm others}) = 0.  \label{eq1-11}
\end{align}
With this commutation relation, as usual, one can interpret $\hat{a}^{\rm in}_{{\bm p},s}$ ($\hat{b}^{{\rm in}\dagger}_{{\bm p},s}$) as an annihilation (creation) operator of one particle (anti-particle) at in-state with quantum number ${\bm p}$ and $s$.

In a similar manner, one can define an annihilation/creation operator at out-state $t \to +\infty$.  To do this, we first define an asymptotic field operator at out-state $\hat{\psi}^{\rm out}$.  Similar to $\hat{\psi}^{\rm in}$, we define an asymptotic field operator at out-state $\hat{\psi}^{\rm out}$ as a solution of the Dirac equation without the weak field ${\mathcal A}_{\mu}$,
\begin{align}
	0 = \left[ i\Slash{\partial} - e\bar{\Slash{A}} -m \right]  \hat{\psi}^{\rm out}, \label{eq1-12}
\end{align}
with a boundary condition at $x^0 \to +\infty$ given by 
\begin{align}
	0 = \lim_{x^0 \to +\infty}\left[ \hat{\psi} - \sqrt{Z} \hat{\psi}^{\rm out}  \right]. \label{eq1-13}
\end{align}
We, then, expand the operator in terms of a mode function ${}_{\pm} \psi^{\rm out}_{{\bm p},s}$ as
\begin{align}
	\hat{\psi}^{\rm out}(x) = \sum_s \int d^3{\bm p} \left[ {}_+ \psi^{\rm out}_{{\bm p},s}(x) \hat{a}^{\rm out}_{{\bm p},s} + {}_- \psi^{\rm out}_{{\bm p},s}(x) \hat{b}^{{\rm out}\dagger}_{-{\bm p},s}   \right],
\end{align}
where we normalize the mode function ${}_{\pm} \psi^{\rm out}_{{\bm p},s}$ in the same manner as ${}_{\pm} \psi^{\rm in}_{{\bm p},s}$ (see Eq.~(\ref{eq1-8})) as
\begin{align}
	\int d^3{\bm x} {}_{\pm} \psi^{{\rm out}\dagger}_{{\bm p},s} {}_{\pm} \psi^{\rm out}_{{\bm p}',s'} &= \delta^3({\bm p}-{\bm p}')\delta_{ss'},\nonumber\\
	\int d^3{\bm x} {}_{\pm} \psi^{{\rm out}\dagger}_{{\bm p},s} {}_{\mp} \psi^{\rm out}_{{\bm p}',s'} &= 0. \label{eq1-15}
\end{align}
The identification of positive/negative frequency mode is essentially the same as ${}_{\pm} \psi^{\rm in}_{{\bm p},s}$ (see Eq.~(\ref{eq1-9})), but is now identified at $x^0 \to +\infty$ as 
\begin{align}
	\lim_{x^0 \to +\infty} {}_{\pm} \psi^{\rm out}_{{\bm p},s} \propto {\rm e}^{\mp i\omega_{{\bm P}^{\rm out}} x^0}{\rm e}^{+i {\bm p}\cdot{\bm x}}, 
\end{align}
where ${\bm P}^{\rm out} \equiv {\bm p}-e\bar{\bm A}(x^0 = +\infty) $ is kinetic momentum at $x^0 \to +\infty$.  Note that ${}_{\pm} \psi^{\rm out}_{{\bm p},s}$ is not necessarily identical to ${}_{\pm} \psi^{\rm in}_{{\bm p},s}$ in the presence of the strong external field.  Since ${}_{\pm} \psi^{\rm out}_{{\bm p},s}$ obeys the same Dirac equation as ${}_{\pm} \psi^{\rm in}_{{\bm p},s}$ does (Eqs.~(\ref{eq1-7}) and (\ref{eq1-12})), ${}_{\pm} \psi^{\rm out}_{{\bm p},s}$ can be written as a superposition of ${}_{+} \psi^{\rm in}_{{\bm p},s}$ and ${}_{-} \psi^{\rm in}_{{\bm p},s}$.  In other words, the positive and negative frequency modes (i.e., particle and anti-particle modes) are mixed up with each other during the time-evolution due to the non-perturbative interactions between $\bar{A}_{\mu}$ and $\hat{\psi}$.  This is one of the important differences from the standard perturbation theory without $\bar{A}_{\mu}$, in which ${}_{\pm} \psi^{\rm in}_{{\bm p},s}$ and ${}_{\pm} \psi^{\rm out}_{{\bm p},s}$ are always identical.  Finally, we impose the canonical commutation relation onto $\hat{\psi}^{\rm out}$, which quantizes $\hat{a}^{\rm out}, \hat{b}^{{\rm out}\dagger}$ as
\begin{align}
	&\delta^3({\bm p}-{\bm p}')\delta_{ss'} = \{ \hat{a}_{{\bm p},s}^{{\rm out}\dagger} ,  \hat{a}_{{\bm p}',s'}^{{\rm out}}  \} = \{ \hat{b}_{{\bm p},s}^{{\rm out}\dagger} ,  \hat{b}_{{\bm p}',s'}^{{\rm out}}  \},  \nonumber\\
	&({\rm others}) = 0.  
\end{align}
Thereby, we define an annihilation/creation operator at out-state.

The annihilation/creation operators at the different asymptotic times, $\hat{a}^{\rm in}, \hat{b}^{{\rm in}\dagger}$ and $\hat{a}^{\rm out}, \hat{b}^{{\rm out}\dagger}$, are not independent with each other.  If the external fields are merely pure gauge fields $\bar{A}_{\mu}, {\mathcal A}_{\mu} \to {\rm const.}$ (i.e., no electromagnetic fields), they are identical.  This is a trivial situation and it is apparent that no particles are produced for this case.  In contrast, for non-vanishing electromagnetic fields $\bar{A}_{\mu} \neq {\rm const.}$ or ${\mathcal A}_{\mu} \neq {\rm const.}$, they are no longer identical but related with each other by a unitary transformation.  We shall see below that this mismatch between the in- and out-state annihilation/creation operators results in particle production.  Note that not only the non-perturbative interactions due to $\bar{A}_{\mu}$, which result in the mixing of the mode functions ${}_{\pm} \psi^{\rm in/out}_{{\bm p},s}$, but also the perturbative interactions due to ${\mathcal A}_{\mu}$ contribute to this mismatch of the annihilation/creation operators.

\subsection{Particle production} \label{sec2-c}

The momentum distribution of produced particles (anti-particles) $n_{{\bm p},s}$ ($\bar{n}_{{\bm p},s}$) can be computed as the in-vacuum expectation value of the number operator at out-state,
\begin{align}
	n_{{\bm p},s}			
		&\equiv \frac{  \braket{{\rm vac;in}| \hat{a}^{{\rm out}\dagger}_{{\bm p},s} \hat{a}^{{\rm out}}_{{\bm p},s}  |{\rm vac;in}}  }{  \braket{{\rm vac;in}|{\rm vac;in}} } ,\nonumber\\
	\bar{n}_{{\bm p},s}
		&\equiv \frac{  \braket{{\rm vac;in}| \hat{b}^{{\rm out}\dagger}_{{\bm p},s} \hat{b}^{{\rm out}}_{{\bm p},s}  |{\rm vac;in}}  }{  \braket{{\rm vac;in}|{\rm vac;in}} } , \label{eq1--18}
\end{align}
where $\ket{{\rm vac;in}}$ is the in-vacuum state, which is a state such that it is annihilated by the annihilation operators at in-state as 
\begin{align}
	0 = \hat{a}^{{\rm in}} \ket{{\rm vac;in}} =  \hat{b}^{{\rm in}} \ket{{\rm vac;in}}.  
\end{align}

We evaluate Eq.~(\ref{eq1--18}) in the lowest non-trivial order of ${\mathcal A}_{\mu}$. To do this, we first rewrite $\hat{a}^{\rm out}, \hat{b}^{{\rm out}\dagger}$ in terms of  $\hat{a}^{\rm in}, \hat{b}^{{\rm in}\dagger}$.  By using Eq.~(\ref{eq1-15}), one can re-express $\hat{a}^{\rm out}, \hat{b}^{{\rm out}\dagger}$ as
\begin{align}
	\begin{pmatrix}
		\hat{a}^{\rm out}_{{\bm p},s} \\ 
		\hat{b}^{{\rm out}\dagger}_{-{\bm p},s} 
	\end{pmatrix}
	&=
	\int d^3{\bm x} 
	\begin{pmatrix}
		{}_{+} \psi^{{\rm out}\dagger}_{{\bm p},s} \\
		{}_{-} \psi^{{\rm out}\dagger}_{{\bm p},s} 
	\end{pmatrix}
	\hat{\psi}^{\rm out} .  
\end{align}
Then, we use the boundary condition (\ref{eq1-13}) and Eq.~(\ref{eq1-4}) to find
\begin{widetext} 
\begin{align}	
	\begin{pmatrix}
		\hat{a}^{\rm out}_{{\bm p},s} \\ 
		\hat{b}^{{\rm out}\dagger}_{-{\bm p},s} 
	\end{pmatrix}
	&=
	\lim_{x^0 \to +\infty} \frac{1}{\sqrt{Z}} \int d^3{\bm x} 
	\begin{pmatrix}
		{}_{+} \psi^{{\rm out}\dagger}_{{\bm p},s}(x) \\
		{}_{-} \psi^{{\rm out}\dagger}_{{\bm p},s}(x) 
	\end{pmatrix}
	\hat{\psi}(x) \nonumber\\
	&= 
	\lim_{x^0 \to +\infty} \int d^3{\bm x} 
	\begin{pmatrix}
		{}_{+} \psi^{{\rm out}\dagger}_{{\bm p},s}(x) \\
		{}_{-} \psi^{{\rm out}\dagger}_{{\bm p},s}(x) 
	\end{pmatrix}
	\left[  
			\hat{\psi}^{\rm in}(x) + e \int d^4 y S_{\rm R}(x,y) \Slash{\mathcal A}(y) \hat{\psi}^{\rm in}(y) + {\mathcal O}(e^2)
	\right]. 	\label{eq1-19}
\end{align}
By noting that the Green function $S_{\rm R}$ can be expressed in terms of the mode function ${}_{\pm} \psi^{{\rm out}}_{{\bm p},s}$ as
\begin{align}
	S_{\rm R}(x,y) = -i\theta(x^0 - y^0) \sum_s \int d^3{\bm p} \left[  {}_{+} \psi^{{\rm out}}_{{\bm p},s}(x) {}_{+} \bar{\psi}^{{\rm out}}_{{\bm p},s}(y)   +  {}_{-} \psi^{{\rm out}}_{{\bm p},s}(x) {}_{-} \bar{\psi}^{{\rm out}}_{{\bm p},s}(y) \right]  , 
\end{align}
one can evaluate Eq.~(\ref{eq1-19}) in the first order of ${\mathcal A}_{\mu}$ as
\begin{align}
	\hat{a}^{\rm out}_{{\bm p},s} 
		= \hat{a}^{{\rm out}(0)}_{{\bm p},s} + \hat{a}^{{\rm out}(1)}_{{\bm p},s} , \ \ 
	\hat{b}^{\rm out}_{-{\bm p},s} 
		= \hat{b}^{{\rm out}(0)\dagger}_{-{\bm p},s} + \hat{b}^{{\rm out}(1)\dagger}_{-{\bm p},s} , \label{eq1-21}
\end{align}
where
\begin{align}
	\hat{a}^{{\rm out}(0)}_{{\bm p},s} &= \sum_{s'} \int d^3{\bm p}' \left[ \left( \int d^3{\bm x} {}_{+} \psi^{{\rm out}\dagger}_{{\bm p},s}(x) {}_+ \psi^{\rm in}_{{\bm p}',s'}(x) \right) \hat{a}^{\rm in}_{{\bm p}',s'} + \left( \int d^3{\bm x} {}_{+} \psi^{{\rm out}\dagger}_{{\bm p},s}(x) {}_- \psi^{\rm in}_{{\bm p}',s'}(x) \right) \hat{b}^{{\rm in}\dagger}_{-{\bm p}',s'} \right], \nonumber\\
	\hat{a}^{{\rm out}(1)}_{{\bm p},s} &=  \sum_{s'} \int d^3{\bm p}' \left[ \left( - ie\int d^4x {}_{+} \bar{\psi}^{{\rm out}}_{{\bm p},s}(x) \Slash{\mathcal A} (x) {}_+ \psi^{{\rm in}}_{{\bm p}',s'}(x) \right) \hat{a}^{{\rm in}}_{{\bm p}',s'} + \left( - ie\int d^4x {}_{+} \bar{\psi}^{{\rm out}}_{{\bm p},s}(x) \Slash{\mathcal A} (x) {}_-\psi^{{\rm in}}_{{\bm p}',s'}(x) \right) \hat{b}^{{\rm in}\dagger}_{-{\bm p}',s'} \right] , 
\end{align}
and
\begin{align}
	\hat{b}^{{\rm out}(0)\dagger}_{-{\bm p},s} &= \sum_{s'} \int d^3{\bm p}' \left[ \left( \int d^3{\bm x} {}_{-} \psi^{{\rm out}\dagger}_{{\bm p},s}(x) {}_+ \psi^{\rm in}_{{\bm p}',s'}(x) \right) \hat{a}^{\rm in}_{{\bm p}',s'} + \left( \int d^3{\bm x} {}_{-} \psi^{{\rm out}\dagger}_{{\bm p},s}(x) {}_- \psi^{\rm in}_{{\bm p}',s'}(x) \right) \hat{b}^{{\rm in}\dagger}_{-{\bm p}',s'} \right], \nonumber\\
	\hat{b}^{{\rm out}(1)\dagger}_{-{\bm p},s} &=  \sum_{s'} \int d^3{\bm p}' \left[ \left( - ie\int d^4x {}_{-} \bar{\psi}^{{\rm out}}_{{\bm p},s}(x) \Slash{\mathcal A} (x) {}_+ \psi^{{\rm in}}_{{\bm p}',s'}(x) \right) \hat{a}^{{\rm in}}_{{\bm p}',s'} + \left( - ie\int d^4x {}_{-} \bar{\psi}^{{\rm out}}_{{\bm p},s}(x) \Slash{\mathcal A} (x) {}_-\psi^{{\rm in}}_{{\bm p}',s'}(x) \right) \hat{b}^{{\rm in}\dagger}_{-{\bm p}',s'} \right] .
\end{align}
An important point here is that, once interactions are switched on, the annihilation operators at out-state $\hat{a}^{{\rm out}}, \hat{b}^{{\rm out}}$ differ from those at in-state $\hat{a}^{{\rm in}}, \hat{b}^{{\rm in}}$ and always contain creation operators at in-state $\hat{a}^{{\rm in}\dagger}, \hat{b}^{{\rm in}\dagger}$.  Hence, the in-vacuum state is no longer annihilated by the annihilation operators at out-state $0 \neq \hat{a}^{{\rm out}} \ket{{\rm vac;in}} , \hat{b}^{{\rm out}} \ket{{\rm vac;in}}$.  This implies that the particle number $n \propto |\hat{a}^{{\rm out}} \ket{{\rm vac;in}}|^2, \bar{n} \propto |\hat{b}^{{\rm out}} \ket{{\rm vac;in}}|^2$ become non-vanishing, i.e., particles are produced from the vacuum.

Now, we can explicitly write down a formula for the particle number.  By substituting Eq.~(\ref{eq1-21}) into Eq.~(\ref{eq1--18}), we obtain
\begin{align}
	n_{{\bm p},s}
		&= \sum_{s'} \int d^3{\bm p}' \left| \int d^3{\bm x} {}_{+} \psi^{{\rm out}\dagger}_{{\bm p},s}(x) {}_- \psi^{\rm in}_{{\bm p}',s'}(x) - ie \int d^4x {}_{+} \bar{\psi}^{{\rm out}}_{{\bm p},s}(x) \Slash{\mathcal A} (x) {}_-\psi^{{\rm in}}_{{\bm p}',s'}(x) \right|^2, \nonumber\\
	\bar{n}_{{\bm p},s}
		&= \sum_{s'} \int d^3{\bm p}' \left| \int d^3{\bm x} {}_{-} \psi^{{\rm out}\dagger}_{-{\bm p},s}(x) {}_+ \psi^{\rm in}_{-{\bm p}',s'}(x) -ie \int d^4x {}_{-} \bar{\psi}^{{\rm out}}_{-{\bm p},s}(x) \Slash{\mathcal A} (x) {}_+\psi^{{\rm in}}_{-{\bm p}',s'}(x)  \right|^2 .  \label{eq2-28}
\end{align}
\end{widetext} 
The first term in the brackets does not contain the weak field ${\mathcal A}_{\mu}$ and it is completely determined by the non-perturbative interactions due to the strong field $\bar{A}_{\mu}$.  Thus, the first term is important for the non-perturbative particle production by the strong field (the Schwinger mechanism).  On the other hand, the second term is important for the perturbative particle production by the weak field.  This is because, for vanishing $\bar{A}_{\mu}$, our formalism reduces to the standard perturbation theory without $\bar{A}_{\mu}$, in which only the second term survives.  However, it should be emphasized that our perturbation theory differs from the standard one because our fermion mode function ${}_{\pm}\psi_{{\bm p},s}^{\rm in/out}$ is fully dressed by the strong field $\bar{A}_{\mu}$.  Thus, the second term depends on $e$ and $\bar{A}_{\mu}$ non-linearly.  Note that our perturbation theory is valid no matter how slow or fast the weak field ${\mathcal A}_{\mu}$ is as long as it is sufficiently weaker than the strong one ${\mathcal A}_{\mu} \ll \bar{A}_{\mu}$.

\section{Constant homogeneous electric field + perturbation} \label{sec3}

In this section, we consider a specific field configuration and discuss details of the particle production based on the perturbation theory developed in Sec.~\ref{sec2}.  In Sec.~\ref{sec3-a}, we first consider a case, in which the external fields $\bar{A}_{\mu}, {\mathcal A}_{\mu}$ are homogeneous in space, and the strong field $\bar{A}_{\mu}$ is sufficiently slow so that it is well approximated by a constant electric field.  For this case, one can analytically perform the integrations in Eq.~(\ref{eq2-28}) without any approximations to obtain a closed expression for the particle number.  This enables us to better understand qualitative aspects of the particle production.  In Sec.~\ref{sec3-b}, we furthermore assume that the weak field is given by a monochromatic wave with frequency $\Omega$, and discuss details of the particle production quantitatively.  In particular, we explicitly demonstrate that the interplay between the non-perturbative particle production (the Schwinger mechanism) and the perturbative one occurs with changing $\Omega$.  Also, we explicitly demonstrate that the Franz-Keldysh effect takes place as a cooperative effect between ${\mathcal A}_{\mu}$ and $\bar{A}_{\mu}$, and it significantly modifies the perturbative particle production.

\subsection{General perturbations} \label{sec3-a}

We assume that the external field is homogeneous in space.  This assumption is equivalent to assume that the external field is purely electric.  By defining the direction of the electric field as the $x^3$-direction, we may write the external fields $\bar{A}_{\mu}, {\mathcal A}_{\mu}$ as
\begin{align}
	\bar{A}_{\mu}(x)		&= (0,0,0, -\bar{A}(x^0)) = (0,0,0,\int^{x^0} dx^0 \bar{E}(x^0))\nonumber\\
	{\mathcal A}_{\mu}(x)	&= (0,0,0,-{\mathcal A}(x^0)) = (0,0,0,\int^{x^0} dx^0 {\mathcal E}(x^0)),
\end{align}
where $\bar{E}, {\mathcal E}$ denote the electric field strength for the strong field and the weak field, respectively.  Furthermore, we assume that $\bar{E}$ is sufficiently slow so that it is well approximated by a constant electric field as
\begin{align}
	\bar{E}(x^0) = \bar{E},\ \ \bar{A}(x^0) = - \bar{E} x^0.  
\end{align}
For simplicity, we assume $\bar{E}>0$ in the following.

For this case, one can analytically solve the Dirac equation, and finds that the mode functions ${}_{\pm} \psi_{{\bm p},s}^{\rm as}$ (${\rm as}={\rm in},{\rm out}$) are given by \cite{nik70, tan09}
\begin{align}
	{}_{+} \psi_{{\bm p},s}^{\rm as}(x) &= \left[ A_{{\bm p}}^{\rm as}(x^0) + B_{{\bm p}}^{\rm as}(x^0) \gamma^0 \frac{ m + {\bm \gamma}_{\perp} \cdot {\bm p}_{\perp}}{\sqrt{m^2 + {\bm p}_{\perp}^2} } \right] \Gamma_s \frac{{\rm e}^{i{\bm p}\cdot {\bm x}}}{(2\pi)^{3/2}} , \nonumber\\
	{}_{-} \psi_{{\bm p},s}^{\rm as}(x) &= \left[ B_{{\bm p}}^{{\rm as}*}(x^0) - A_{{\bm p}}^{{\rm as}*}(x^0) \gamma^0 \frac{ m + {\bm \gamma}_{\perp} \cdot {\bm p}_{\perp}}{\sqrt{m^2 + {\bm p}_{\perp}^2} } \right] \Gamma_s \frac{{\rm e}^{i{\bm p}\cdot {\bm x}}}{(2\pi)^{3/2}} .  \label{eq3-31}
\end{align}
Here, $\Gamma_s$ ($s = \uparrow, \downarrow$) are two eigenvectors of $\gamma^0 \gamma^3$ with eigenvalue one such that
\begin{align}
	\gamma^0 \gamma^3 \Gamma_s = \Gamma_s,\ \ \Gamma_s^{\dagger}\Gamma_{s'} = \delta_{ss'}, 
\end{align}
and the scalar functions $ A_{{\bm p}}^{\rm as},  B_{{\bm p}}^{\rm as}$ are 
\begin{align}
	&\!\!\left\{\begin{array}{l}
		\!A^{\rm in}_{{\bm p}} \!=\!  {\rm e}^{-\frac{i\pi}{8}} {\rm e}^{-\frac{\pi}{8}  \!\frac{m_{\perp}^2}{e\bar{E}}}\! \frac{m_{\perp}}{\sqrt{2e\bar{E}}} D_{\!\frac{i}{2}\! \frac{m_{\perp}^2}{e\bar{E}}-1} \!\!\left(-{\rm e}^{-\frac{i\pi}{4}} \sqrt{\frac{2}{e\bar{E}}} (e\bar{E}x^0 + p_{\parallel}) \right) \\
		\!B^{\rm in}_{{\bm p}} \!=\!  {\rm e}^{+\frac{i\pi}{8}} {\rm e}^{-\frac{\pi}{8}\! \frac{m_{\perp}^2 }{e\bar{E}}} D_{\!\frac{i}{2} \!\frac{m_{\perp}^2}{e\bar{E}}} \!\!\left(-{\rm e}^{-\frac{i\pi}{4}} \sqrt{\frac{2}{e\bar{E}}} (e\bar{E}x^0 + p_{\parallel}) \right)
	\end{array}\right., \nonumber\\
	&\!\!\left\{\begin{array}{l}
		\!A^{\rm out}_{{\bm p}} \!=\!  {\rm e}^{-\frac{i\pi}{8}} {\rm e}^{-\frac{\pi}{8}\! \frac{m_{\perp}^2 }{e\bar{E}}}  D_{\!-\frac{i}{2} \!\frac{m_{\perp}^2}{e\bar{E}}} \!\!\left({\rm e}^{\frac{i\pi}{4}} \sqrt{\frac{2}{e\bar{E}}} (e\bar{E}x^0 + p_{\parallel}) \right) \\
		\!B^{\rm out}_{{\bm p}} \!=\!  {\rm e}^{+\frac{i\pi}{8}} {\rm e}^{-\frac{\pi}{8}\! \frac{m_{\perp}^2}{e\bar{E}}} \! \frac{m_{\perp}}{\sqrt{2e\bar{E}}}  D_{\!-\frac{i}{2}\! \frac{m_{\perp}^2}{e\bar{E}}-1} \!\!\left({\rm e}^{\frac{i\pi}{4}} \sqrt{\frac{2}{e\bar{E}}} (e\bar{E}x^0 + p_{\parallel}) \right)
	\end{array}\right. ,  \label{eq3-33}
\end{align}
where $D_{\nu}(z)$ is the parabolic cylinder function\footnote{In non-relativistic systems, the mode function in the presence of a constant electric field is expressed by the Airy function, not by the parabolic cylinder function.  This is a slight difference between the Franz-Keldysh effect in QED and in semi-conductors.  }, and 
\begin{align}
	m_{\perp} \equiv \sqrt{m^2 + {\bm p}_{\perp}^2}
\end{align}
is transverse mass.  $p_{\parallel}, {\bm p}_{\perp}$ are longitudinal and transverse momentum with respect to the direction of the electric field, respectively.  Note that ${}_{\pm} \psi^{\rm in}_{{\bm p},s}$ and ${}_{\pm} \psi^{\rm out}_{{\bm p},s}$ are not linearly independent with each other, but are related with each other by
\begin{align}
	\begin{pmatrix}
		{}_{+} \psi^{\rm in}_{{\bm p},s} \\
		{}_{-} \psi^{\rm in}_{{\bm p},s} 
	\end{pmatrix}
	=
	\begin{pmatrix}
		\alpha_{{\bm p}} & -\beta^*_{{\bm p}}\\
		\beta_{{\bm p}} & \alpha^*_{{\bm p}}
	\end{pmatrix}
	\begin{pmatrix}
		{}_{+} \psi^{\rm out}_{{\bm p},s} \\
		{}_{-} \psi^{\rm out}_{{\bm p},s} 
	\end{pmatrix} , 
\end{align}
where
\begin{align}
	\alpha_{{\bm p}} &\equiv \frac{m_{\perp}}{\sqrt{2e\bar{E}}} \frac{\sqrt{2\pi} \exp\left[ - \frac{\pi}{4} \frac{m^2_{\perp}}{e\bar{E}} \right]}{ \Gamma\left(1- \frac{i}{2}\frac{m^2_{\perp}}{e\bar{E}} \right) },\nonumber\\ 
	\beta_{{\bm p}} &\equiv \exp\left[ - \frac{\pi}{2} \frac{m^2_{\perp}}{e\bar{E}} \right] .  
\end{align}
It should be stressed that $|\alpha_{\bm p}| \neq 1$ and $|\beta_{\bm p}| \neq 0$ if $\bar{E} \neq 0$.  $\alpha_{\bm p}, \beta_{\bm p}$ can be understood as an analog of the reflectance and the transmission coefficient in a barrier scattering problem in quantum mechanics, respectively.  Thus, intuitively speaking, $|\alpha_{\bm p}| \neq 1$ ($|\beta_{\bm p}| \neq 0$) implies that particles in the Dirac sea are reflected by (tunneled into) the tilted gap in the presence of the strong electric field.  This point plays an important role in the appearance of the Franz-Keldysh effect as we shall explain later\footnote{In general, if there exists a ``genuine'' strong electric field which cannot be eliminated by any Lorentz transformations, $|\alpha_{\bm p}| \neq 1$ and $|\beta_{\bm p}| \neq 0$ hold.  This implies that the Franz-Keldysh effect is a genuinely electrical effect.  For example, strong plane waves, strong crossed fields, or strong magnetic field alone always gives $|\alpha_{\bm p}| = 1$ and $|\beta_{\bm p}| = 0$ no matter how strong it is, and hence the Franz-Keldysh effect never occurs.  }.

With the use of Eqs.~(\ref{eq3-31}) and (\ref{eq3-33}), one can evaluate the integrals in Eq.~(\ref{eq2-28}) analytically as
\begin{align}
	&\int d^3{\bm x} {}_{-} \psi^{{\rm out}\dagger}_{{\bm p},s} {}_{+} \psi^{\rm in}_{{\bm p}',s'} \nonumber\\
	&= - \left(  \int d^3{\bm x} {}_{+} \psi^{{\rm out}\dagger}_{{\bm p},s} {}_{-} \psi^{\rm in}_{{\bm p}',s'} \right)^* \nonumber\\
	&= \delta_{ss'}\delta^3({\bm p}-{\bm p}') \times \exp\left[ - \frac{\pi}{2} \frac{m^2_{\perp}}{e\bar{E}} \right] , 
\end{align}
and
\begin{align}
	& -ie \int d^4x {}_{-} \bar{\psi}^{{\rm out}}_{{\bm p},s} \Slash{\mathcal A} {}_{+} \psi^{{\rm in}}_{{\bm p}',s'} \nonumber\\
	&= - \left( -ie \int d^4x {}_{+} \bar{\psi}^{{\rm out}}_{{\bm p},s} \Slash{\mathcal A} {}_{-} \psi^{{\rm in}}_{{\bm p}',s'} \right)^* \nonumber\\
	&= \delta_{ss'}\delta^3({\bm p}-{\bm p}') \times \frac{1}{2} \frac{m_{\perp}^2}{e\bar{E}} \exp\left[ - \frac{\pi}{2} \frac{m^2_{\perp}}{e\bar{E}} \right] \nonumber\\
	&\quad \times\int_0^{\infty} d\omega \frac{\tilde{\mathcal E}(\omega)}{\bar{E}} \exp\left[ - \frac{i}{4}\frac{\omega^2 + 4\omega p_z}{e\bar{E}}  \right] \nonumber\\
	&\quad \times {}_1\tilde{F}_1 \left( 1-\frac{i}{2}\frac{m_{\perp}^2}{e\bar{E}}; 2; \frac{i}{2} \frac{\omega^2}{e\bar{E}} \right).  
\end{align}
Here, ${}_1\tilde{F}_1$ is the regularized hypergeometric function, and we introduced the Fourier transformation of the weak electric field as 
\begin{align}
	\tilde{\mathcal E}(\omega) \equiv \int dx^0 {\rm e}^{-i\omega x^0} {\mathcal E}(x^0).  
\end{align}
Note that the off-diagonal matrix elements with different momentum ${\bm p}\neq{\bm p}'$ in the integrals are vanishing because of the spatial homogeneity.  Different spin labels $s \neq s'$ also give vanishing contributions because there are no magnetic fields and electric fields do not couple to spins.  By plugging these expressions into Eq.~(\ref{eq2-28}), one finally obtains
\begin{widetext}
\begin{align}
	n_{{\bm p},s} = \bar{n}_{-{\bm p},s}
	= \frac{V}{(2\pi)^3}  \exp\left[ -\pi \frac{m^2_{\perp}}{e\bar{E}} \right] \left| 1 +  \frac{1}{2} \frac{m^2_{\perp}}{e\bar{E}} \int_0^{\infty} d\omega \frac{\tilde{\mathcal E}(\omega)}{\bar{E}} \exp\left[ -\frac{i}{4}\frac{\omega^2 + 4\omega p_{\parallel}}{e\bar{E}}  \right] {}_1\tilde{F}_1 \left( 1- \frac{i}{2} \frac{m^2_{\perp}}{e\bar{E}}; 2; \frac{i}{2} \frac{\omega^2}{e\bar{E}} \right) \right|^2, \label{eq3-40}
\end{align}
\end{widetext}
where the use is made of $\delta^3({\bm p}={\bm 0}) = V/(2\pi)^3$ with $V$ being the whole spatial volume.  Equation (\ref{eq3-40}) does not depend on $s$ because electric fields do not distinguish spins.  $n_{{\bm p}} = \bar{n}_{-{\bm p}}$ holds because a particle and an anti-particle are always produced together as a pair from the vacuum, whose momentum and charge are zero.  Note that we did not use any approximations (such as the WKB approximation) in deriving Eq.~(\ref{eq3-40}).

\subsubsection{Non-perturbative limit} 

The particle production becomes non-perturbative (i.e., the Schwinger mechanism occurs) if the weak electric field ${\mathcal E}$ is so slow that $\tilde{\mathcal E}$ is dominated by low-frequency modes $\omega/\sqrt{e\bar{E}} \ll 1$.

Indeed, by taking a limit of $\omega/\sqrt{e\bar{E}} \to 0$ in the integrand of Eq.~(\ref{eq3-40}), we obtain
\begin{align}
	&n_{{\bm p},s} = n_{-{\bm p},s} \nonumber\\
	&\sim \frac{V}{(2\pi)^3} \exp\left[  -\pi \frac{m_{\perp}^2}{e \bar{E}}  \right] \left| 1 + \frac{1}{2} \frac{m_{\perp}^2}{e\bar{E}} \int^{\infty}_0 d\omega {\rm e}^{-i \omega \frac{p_{\parallel}}{e\bar{E}} } \frac{\tilde{\mathcal E}(\omega)}{\bar{E}} \right|^2 \nonumber\\
	&\sim \frac{V}{(2\pi)^3} \exp\left[  -\pi \frac{m_{\perp}^2}{e \bar{E}}  \right] \left| 1 + \frac{\pi}{2} \frac{m_{\perp}^2}{e\bar{E}}  \frac{{\mathcal E}(-p_{\parallel}/e\bar{E})}{\bar{E}} \right|^2.     \label{eq3-42}  
\end{align}
In the last line, we used a mathematical trick $\int_0^{\infty} d\omega {\rm e}^{-i \omega t} \sim \pi \delta(t)$.  In Eq.~(\ref{eq3-42}), the coupling constant $e$ appears inversely in the exponential.  This fact ensures that the particle production is actually non-perturbative for slow ${\mathcal E}$.  Note that the distribution depends on $p_{\parallel}$ if ${\mathcal E}$ depends on time.  Intuitively, the particle production occurs most efficiently at the instant when the longitudinal kinetic momentum $P_{\parallel} = p_{\parallel} + e\bar{E} x^0$ becomes zero, at which the energy cost to produce a particle is the smallest.  Thus, the value of the weak field at $x^0 = -p_{\parallel}/e\bar{E}$ becomes important.

Equation~(\ref{eq3-42}) is consistent with the Schwinger formula for the non-perturbative particle production from a constant electric field.  In fact, the Schwinger formula reads \cite{sch51}
\begin{align}
	&n^{\rm (Sch)}_{{\bm p},s} = \bar{n}^{\rm (Sch)}_{-{\bm p},s} \nonumber\\
	&= \frac{V}{(2\pi)^3} \exp\left[  -\pi \frac{m_{\perp}^2}{eE}  \right]  \nonumber\\
	&= \frac{V}{(2\pi)^3} \exp\left[  -\pi \frac{m_{\perp}^2}{e\bar{E}}  \right] \left| 1 + \frac{\pi}{2} \frac{m_{\perp}^2}{e\bar{E}} \frac{\mathcal E}{\bar{E}} + {\mathcal O} \left( \left( \frac{\mathcal E}{\bar{E}}\right)^2 \right) \right|^2 , \label{eq3-43}
\end{align}
where $E = \bar{E} + {\mathcal E}$ is the total electric field strength.  Thus, our formula (\ref{eq3-42}) reproduces the Schwinger formula (\ref{eq3-43}) up to ${\mathcal O} ( ( {\mathcal E}/\bar{E} )^1 )$ if one regards ${\mathcal E}(-p_{\parallel}/g\bar{E})$ as a constant.  To reproduce ${\mathcal O} (( {\mathcal E}/\bar{E} )^n )$-corrections ($n \geq 2$) correctly within our perturbation theory, one has to expand the annihilation operators (\ref{eq1-19}) up to $n$-th order in ${\mathcal A}_{\mu}$.

\subsubsection{Perturbative limit}

The perturbative particle production takes place if the weak electric field $\tilde{\mathcal E}$ is dominated by high-frequency modes.  Indeed, by taking $\omega/\sqrt{e\bar{E}} \to \infty$ limit of the integrand in Eq.~(\ref{eq3-40}), we obtain
\begin{align}
	n_{{\bm p},s} = \bar{n}_{-{\bm p},s} \sim \frac{V}{(2\pi)^3}  \left| \exp\left[ - \frac{\pi}{2} \frac{m^2_{\perp}}{e\bar{E}} \right] +  \frac{1}{2} \frac{m_{\perp}}{\omega_{\bm p}}\frac{ e\tilde{\mathcal E}(2\omega_{\bm p}) }{\omega_{\bm p}} \right|^2.  \label{eq3-45}
\end{align}
Equation (\ref{eq3-45}) is a superposition of the non-perturbative and the perturbative particle production.  In fact, the second term does not contain the exponential factor, but just depends on $e$ linearly.  Hence, it gives the perturbative particle production.  The perturbative particle production does not depend on $\bar{E}$, but solely determined by ${\mathcal E}$.  The strong field $\bar{E}$ separately contributes to the scattering amplitude, and gives rise to the non-perturbative particle production.  This is the first term in Eq.~(\ref{eq3-45}), which is independent of ${\mathcal E}$.

If $\bar{E}$ is smaller than the critical field strength $e\bar{E} \lesssim m_{\perp}^2$, the first term in Eq.~(\ref{eq3-45}) may be neglected because it is exponentially suppressed.  Thus, Eq.~(\ref{eq3-45}) becomes purely perturbative as 
\begin{align}
	n_{{\bm p},s} = \bar{n}_{-{\bm p},s} \sim \frac{V}{(2\pi)^3} \frac{1}{4} \frac{m^2_{\perp}}{\omega^2_{\bm p}}\frac{ \left| e\tilde{\mathcal E}(2\omega_{\bm p}) \right|^2 }{\omega_{\bm p}^2}  .  \label{eq3-46}
\end{align}
Note that Eq.~(\ref{eq3-46}) reproduces the textbook formula for the perturbative particle production from a classical electric field \cite{itz80, tay14}.

On the other hand, if $\bar{E}$ is super-critical $e\bar{E} \gtrsim m_{\perp}^2$, the first term in Eq.~(\ref{eq3-46}) becomes ${\mathcal O}(1)$, which is superior to the second term ${\mathcal O}(e{\mathcal E}/\omega_{\bm p}^2)$.  Then, Eq.~(\ref{eq3-46}) gives
\begin{align}
	n_{{\bm p},s} = \bar{n}_{-{\bm p},s} \sim \frac{V}{(2\pi)^3} \exp\left[ - \pi \frac{m^2_{\perp}}{e\bar{E}} \right] . 
\end{align}
This implies that the perturbative particle production by ${\mathcal E}$ is buried in the non-perturbative one by $\bar{E}$, and the particle production always looks non-perturbative no matter how slow or fast the weak field is.  In other words, the interplay between the perturbative and the non-perturbative particle production becomes less manifest if $\bar{E}$ is super-critical.

\subsection{Monochromatic wave as a perturbation} \label{sec3-b}

In this section, we consider an explicit example, in which the weak field is given by a monochromatic wave 
\begin{align}
	{\mathcal E}(x^0) = {\mathcal E}_0 \cos \Omega x^0 .  \label{eq3-47}
\end{align}
With this configuration, we compute the momentum distribution $n_{{\bm p},s}$ and the total particle number $N \equiv \sum_s \int d^3{\bm p} \; n_{{\bm p},s}$ to explicitly demonstrate how the interplay between the non-perturbative and the perturbative particle production occurs with changing the frequency $\Omega$.  We also demonstrate that a QED analog of the Franz-Keldysh effect occurs.  The Franz-Keldysh effect significantly lowers the threshold frequency for the perturbative particle production, and results in a characteristic oscillating pattern in $\Omega$-dependence.

\subsubsection{momentum distribution} \label{sec4-b1}

By noting that the Fourier component $\tilde{\mathcal E}$ is sharply peaked at $\omega = \pm |\Omega|$ as
\begin{align}
	\tilde{\mathcal E}(\omega) = \pi {\mathcal E}_0 \left[ \delta(\omega - |\Omega|) + \delta(\omega + |\Omega|)    \right], 
\end{align}
the formula for the number distribution (\ref{eq3-40}) can be simplified as
\begin{align}
	n_{{\bm p},s} = \bar{n}_{-{\bm p},s}
	&= \frac{V}{(2\pi)^3}  \exp\left[ -\pi \frac{m^2_{\perp}}{e\bar{E}} \right] \left| 1 +  \frac{\pi}{2} \frac{m^2_{\perp}}{e\bar{E}} \frac{{\mathcal E}_0}{\bar{E}} \right.\nonumber\\
	&\quad\times \exp\left[ -\frac{i}{4}\frac{|\Omega|^2 + 4|\Omega| p_z}{e\bar{E}}  \right] \nonumber\\
	&\quad\times \left. {}_1\tilde{F}_1 \left( 1- \frac{i}{2} \frac{m^2_{\perp}}{e\bar{E}}; 2; \frac{i}{2} \frac{|\Omega|^2}{e\bar{E}} \right) \right|^2.   \label{eq4-50}
\end{align}

\begin{figure}[!t]
\includegraphics[clip, width=0.499\textwidth]{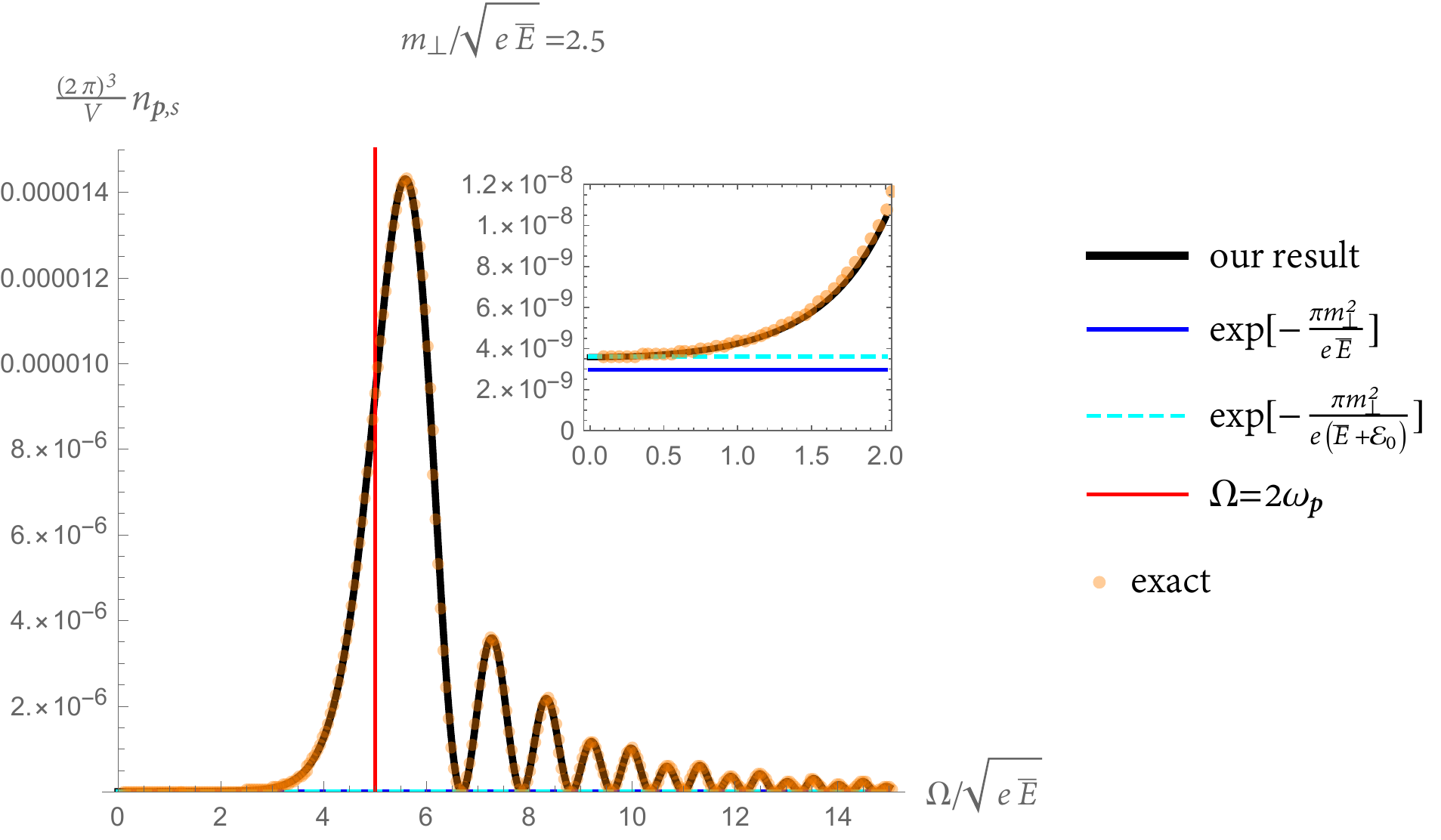}
\includegraphics[clip, width=0.499\textwidth]{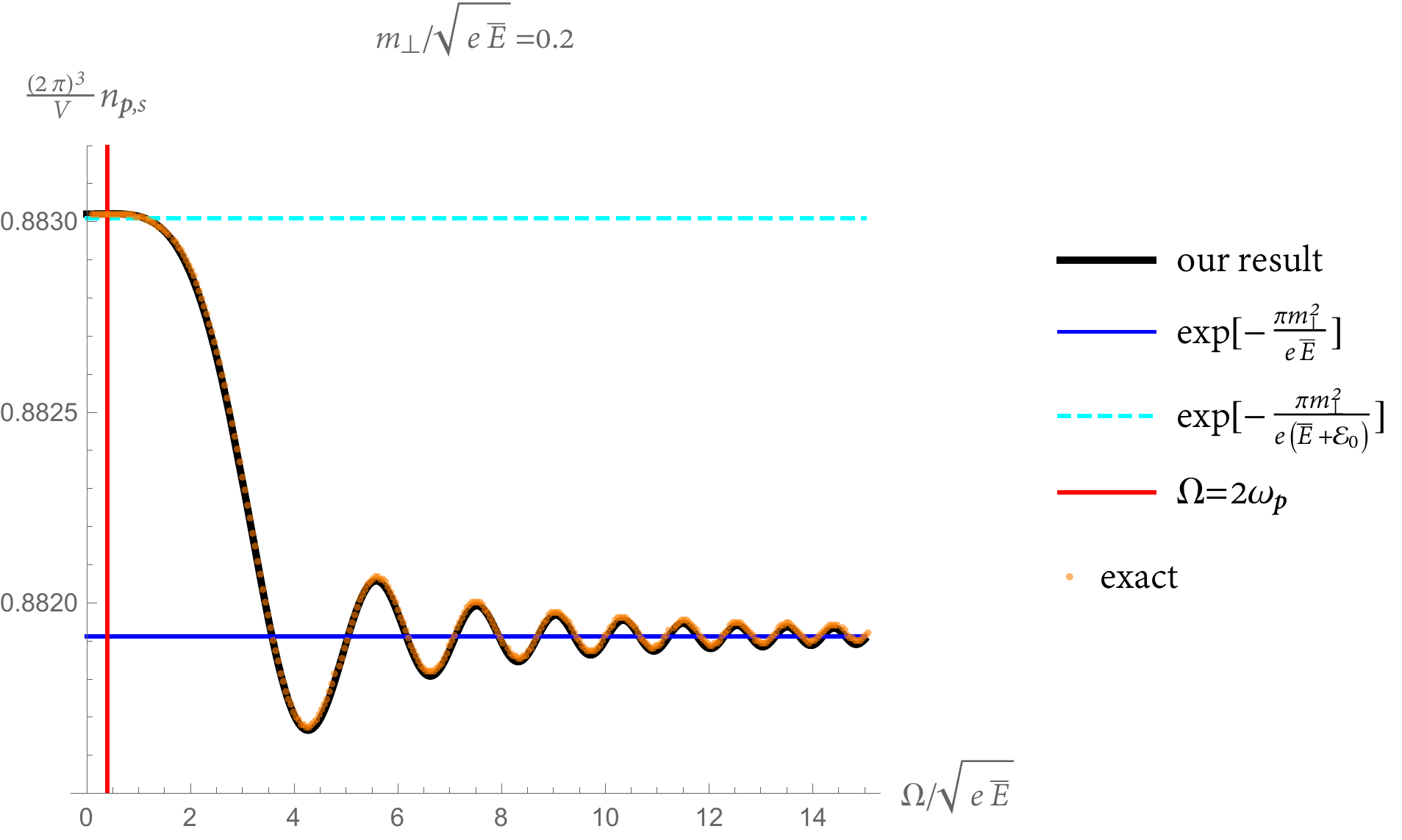}
\caption{\label{fig1} (color online) The frequency $\Omega$-dependence of the number distribution $n_{{\bm p},s}$ (thick black line) for sub-critical field strength $m_{\perp}/\sqrt{e\bar{E}} = 2.5$ (top) and super-critical one $m_{\perp}/\sqrt{e\bar{E}} = 0.2$ (bottom).   We fixed $p_{\parallel}$ and ${\mathcal E}_0$ as $p_{\parallel}/\sqrt{e\bar{E}} = 0$ and ${\mathcal E}_0 = \bar{E}/100$.  The blue and dashed cyan lines show the Schwinger formula (see Eq.~(\ref{eq3-43})) for the strong field alone $\exp[-\pi m_{\perp}^2/e\bar{E}]$ and for the total electric field $\exp[-\pi m_{\perp}^2/e(\bar{E}+{\mathcal E}_0)]$, respectively.  The red vertical line shows $\Omega = 2\omega_{\bm p}$, at which the perturbative particle production is peaked (see Eq.~(\ref{eq3--50})).  The orange dotted line shows the exact result, which is obtained by numerically solving the original Dirac equation (\ref{eq1-2}).}
\end{figure}

\begin{figure}[!t]
\includegraphics[clip, width=0.499\textwidth]{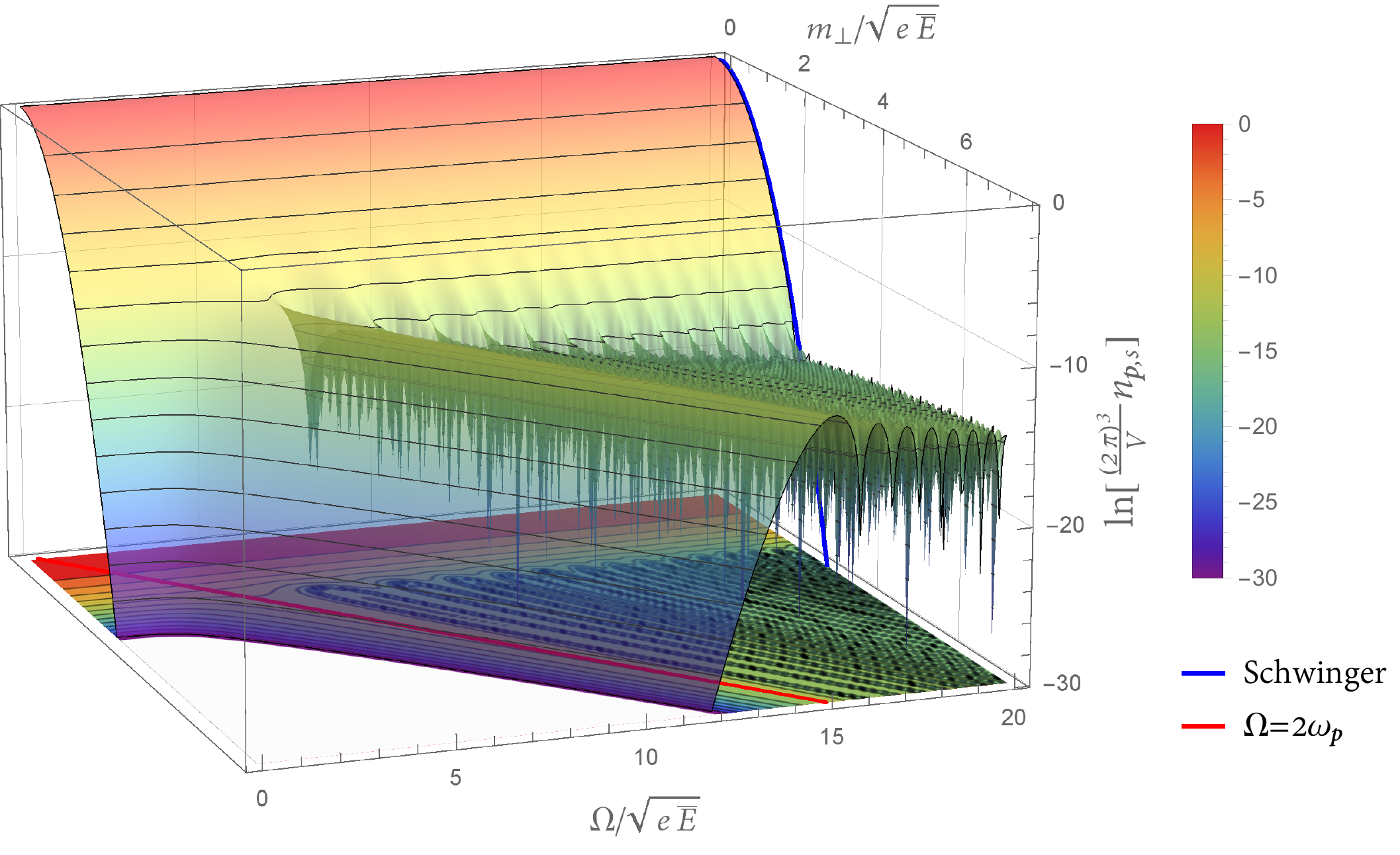}
\caption{\label{fig2} (color online) The number distribution $n_{{\bm p},s}$ as a function of the frequency $\Omega$ and the transverse mass $m_{\perp}$.  $p_{\parallel}$ and ${\mathcal E}_0$ are fixed as $p_{\parallel}/\sqrt{e\bar{E}} = 0$ and ${\mathcal E}_0 = \bar{E}/100$.  The blue line at $\Omega/\sqrt{e\bar{E}} = 20$ shows the Schwinger formula (see Eq.~(\ref{eq3-43})) for the non-perturbative field alone $\exp[-\pi m_{\perp}^2/e\bar{E}]$.  The red line at the bottom shows $\Omega = 2\omega_{\bm p}$, at which the perturbative particle production is peaked (see Eq.~(\ref{eq3--50})).  }
\end{figure}

The number distribution (\ref{eq4-50}) is plotted in Figs.~\ref{fig1} and \ref{fig2}.  We also compared Eq.~(\ref{eq4-50}) with various other evaluations, i.e., the Schwinger non-perturbative formula (\ref{eq3-43}); the perturbative formula (\ref{eq3-46}); and an exact result which is obtained by numerically solving the original Dirac equation (\ref{eq1-2}) without any expansion nor approximations\footnote{We solved the Dirac equation numerically for a finite time interval $x^0 \in [-T,T]$ with a plane wave initial condition set at $x^0 = -T$, and computed the off-diagonal Bogoliubov coefficient $\beta_{\bm p}$ at $x^0 = T$.  The number distribution at time $T$ is obtained as $n_{{\bm p},s} = V/(2\pi)^3 \times |\beta_{\bm p}|^2$ \cite{tan09}.  The exact result plotted in Fig.~\ref{fig1} is obtained by taking sufficiently large $T$.  We carefully checked that the result is insensitive to $T$ if it is large enough.  }.  Notice that Eq.~(\ref{eq4-50}) reproduces the exact result very well for any values of $\Omega$.  This confirms that our perturbative formulation is valid as long as the weak field ${\mathcal E}$ is weak enough ${\mathcal E} \ll \bar{E}$, and that the frequency of ${\mathcal E}$ is not important.

For sub-critical field strength $e\bar{E} \lesssim m_{\perp}^2$, the interplay between the non-perturbative and the perturbative particle production takes place (see the top panel of Fig.~\ref{fig1} and Fig.~\ref{fig2}) as we already discussed analytically in Sec.~\ref{sec3-a}: In the high-frequency region $\Omega/\sqrt{e\bar{E}} \gg 1$, the particle production becomes the most efficient at $\Omega \sim 2\omega_{\bm p}$.  This implies that the production is dominated by the perturbative process.  Indeed, the perturbative formula (\ref{eq3-46}) for the monochromatic wave (\ref{eq3-47}) is sharply peaked at $\Omega = 2\omega_{\bm p}$ as
\begin{align}
	n^{\rm (pert)}_{{\bm p},s} = \frac{VT}{(2\pi)^3} \frac{\pi}{8} \frac{m^2_{\perp}}{\omega^2_{\bm p}}\frac{ \left| e{\mathcal E}_0 \right|^2 }{\omega_{\bm p}^2} \delta(|\Omega|-2\omega_{\bm p})  , \label{eq3--50} 
\end{align}
where we used $\delta(\omega=0)=T/2\pi$ with $T$ being the whole time interval.  Physically speaking, the location of the peak $\Omega = 2 \omega_{\bm p}$ in Eq.~(\ref{eq3--50}) can be understood as the threshold energy for one photon to create a pair of particles from the vacuum.  On the other hand, in the low-frequency region $\Omega/\sqrt{e\bar{E}} \ll 1$, the particle production becomes non-perturbative, and is consistent with the Schwinger formula for the total electric field (\ref{eq3-43}).  Notice that the non-perturbative particle production is strongly suppressed by an exponential of $|e\bar{E}|^{-1}$, but the perturbative one is only suppressed by powers of $e\bar{E}$.  Thus, the perturbative particle production is more abundant than the non-perturbative one for sub-critical field strength $e\bar{E} \lesssim m_{\perp}^2$.

\begin{figure}[!t]
\includegraphics[clip, width=0.45\textwidth]{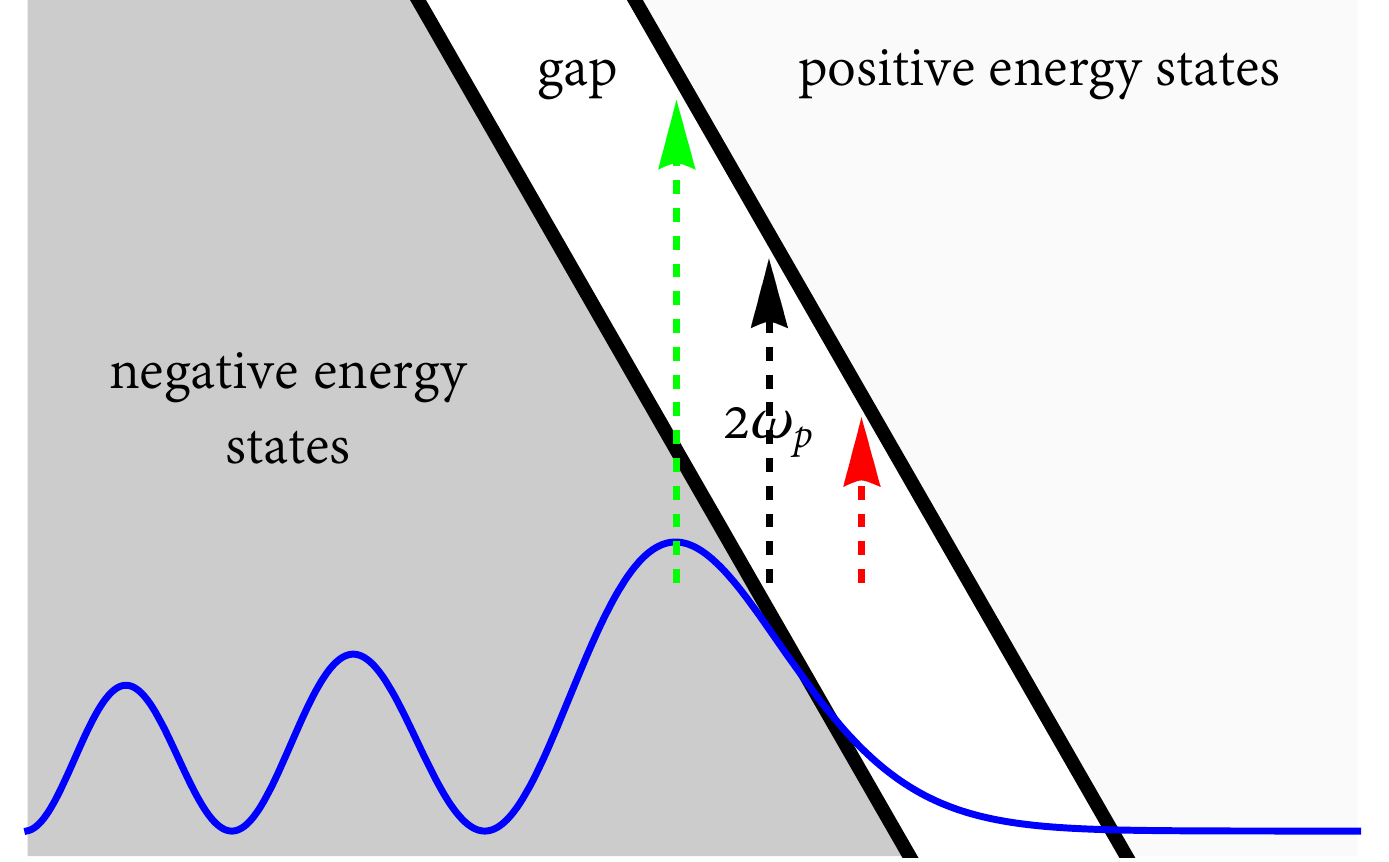}
\caption{\label{fig3} (color online) A schematic picture of the band structure of QED in the presence of a strong constant electric field $\bar{E}$.  The blue curve represents the probability density $\psi^{\dagger} \psi$ of a particle in the Dirac sea.  The black dashed line shows the band gap energy $2 \omega_{\bm p}$ between the band edges.  The red (green) dashed line shows the energy needed to excite a particle in the Dirac sea which is tunneled into (reflected by) the tilted gap into a positive energy state.  }
\end{figure}

The structure of the perturbative peak at $\Omega \sim 2\omega_{\bm p}$ (see the top panel of Fig.~\ref{fig1} and Fig.~\ref{fig2}) is significantly modified by the strong field $\bar{E}$.  This is nothing but the Franz-Keldysh effect in QED.  Indeed, in contrast to the naive perturbative formula (\ref{eq3--50}), the perturbative peak in the figures is not a simple delta function strictly localized at the threshold energy, but it has non-trivial structure: (i) there is a tail below the threshold $\Omega \lesssim 2\omega_{\bm p}$; (ii) the largest peak is located slightly above the threshold $\Omega \gtrsim 2\omega_{\bm p}$; and (iii) above the threshold $\Omega > 2\omega_{\bm p}$, the peak does not decrease monotonically but oscillates.

Here is an intuitive explanation why the Franz-Keldysh effect occurs (see also Fig.~\ref{fig3}): The perturbative particle production from the vacuum occurs when a particle which is filling one of the negative energy states (the Dirac sea) is excited into one of the positive energy states.  As the energy bands are non-perturbatively tilted by the strong electric field $\bar{E}$, the probability density of a particle in the Dirac sea can tunnel into the gap.  Once the particle tunneled into the gap, the threshold energy to excite the particle into a positive state is reduced.  Thus, the perturbative particle production can occur even below the naive threshold (i).  However, this does not necessarily imply that the particle production occurs most efficiently below the threshold.  On the contrary, it should be slightly above the threshold in the presence of $\bar{E}$.  This is because only a part of the probability density can tunnel into the gap but the major part of it is reflected by the gap.  Because of this reflection, the probability density takes its maximum slightly away from the gap, at which more energy is needed to excite the particle.  Hence, the perturbative particle production becomes the most efficient slightly above the naive threshold energy (ii).  Another important consequence of the reflection is that it mixes up the in-coming and out-going wave.  Therefore, the probability density outside of the gap is no longer uniform but oscillates in space.  This results in the oscillating pattern in the distribution (iii) because the excitation energy at each local maximum of the probability density is different and larger excitation energy is needed for deeper local maxima.

For super-critical field strength $e\bar{E} \gtrsim m_{\perp}^2$ (see the bottom panel of Fig.~\ref{fig1} and Fig.~\ref{fig2}), the non-perturbative production becomes so abundant no matter how slow or fast the frequency $\Omega$ is.  Thus, the perturbative production is always buried in the non-perturbative one, and the interplay or the Franz-Keldysh effect is not manifest at first sight.  This, however, does not imply that there is no interplay nor the Franz-Keldysh effect.  Indeed, the distribution shows an oscillating behavior for large $\Omega$, which is a reminiscent of the Franz-Keldysh oscillation (iii).  Also, the production becomes smaller for large $\Omega$, which is because the interplay takes place.  For large $\Omega$, the weak field ${\mathcal E}$ and the strong field $\bar{E}$ separately contribute to the scattering amplitude of the production process.  As the weak field ${\mathcal E}$ with large $\Omega$ only gives a perturbative contribution, which is negligible to the non-perturbative one from $\bar{E}$, the distribution is described well by the Schwinger formula for the strong field $\bar{E}$ alone.  On the other hand, for small $\Omega$, not only $\bar{E}$ but also ${\mathcal E}$ contributes to the production process in a non-perturbative manner.  Thus, the distribution is described by the Schwinger formula for the total field $E = \bar{E} + {\mathcal E}_0$, which gives larger (if ${\mathcal E}_0>0$) production compared to that for $\bar{E}$ alone.

Let us discuss a relation between the Franz-Keldysh effect in QED and the dynamically assisted Schwinger mechanism.  The physical set-ups in the Franz-Keldysh effect and in the dynamically assisted Schwinger mechanism are quite similar to each other.  Although both mechanisms consider a superposition of a weak field onto a strong slow field, it is typical to assume in the dynamically assisted Schwinger that the frequency of the weak field is well below the threshold and the strong electric field is sub-critical.  Thus, the set-up discussed in (i) (i.e., a weak field with frequency below the threshold on top of a sub-critical strong field) is the completely same set-up that is discussed in the dynamically assisted Schwinger mechanism.  Therefore, the physical origin of (i) should be the same as the dynamically assisted Schwinger mechanism.  Indeed, the dynamically assisted Schwinger mechanism claims that the quantum tunneling is assisted by the perturbative excitation, while the pertubative excitation is assisted by the quantum tunneling in the Franz-Keldysh effect.  This is just a rephrasing of the same physical process from a different point of view.  Note that the dominance of the perturbative particle production in the dynamically assisted Schwinger mechanism was emphasized previously in Refs.~\cite{tan16, tor17}.  On the other hand, the set-ups discussed in (ii), (iii) (i.e., a weak field with frequency near and/or above the threshold on top of a sub-critical strong field), and the super-critical set-up are not the set-ups discussed in the context of the dynamically assisted Schwinger mechanism.  Therefore, one may understand the Franz-Keldysh effect in QED as a generalization of the dynamically assisted Schwinger mechanism to broader parameter regions.

Note that a similar effect to (ii) was found previously in Ref.~\cite{koh14}, in which electron and positron pair production from a strong oscillating electric field was discussed.  Ref.~\cite{koh14} found that pair production thresholds for multi-photon processes increase as the strength of the strong oscillating field increases, although Ref.~\cite{koh14} did not consider static strong electric field and interpreted it in terms of a change of electron's effective mass.

\subsubsection{total number} 

The total number of produced particles $N$ can be computed by integrating the spin $s$ and the momentum ${\bm p}$ of the distribution $n_{{\bm p},s}$ (\ref{eq4-50}) as
\begin{align}
	N	&= \sum_s \int d^3{\bm p}\; n_{{\bm p},s} \nonumber\\
		&=  (e\bar{E})^2 VT \times \frac{1}{e\bar{E}} \frac{1}{2 \pi^2} \int_m^{\infty} dm_{\perp} m_{\perp}  \exp\left[ -\pi \frac{m^2_{\perp}}{e\bar{E}} \right] \nonumber\\
	&\quad\times \left[ 1 +  \left|  \frac{\pi}{2}  \frac{m^2_{\perp}}{e\bar{E}} {}_1\tilde{F}_1 \left( 1- \frac{i}{2} \frac{m^2_{\perp}}{e\bar{E}}; 2; \frac{i}{2} \frac{|\Omega|^2}{e\bar{E}} \right) \right|^2 \left(  \frac{{\mathcal E}_0}{\bar{E}}  \right)^2\right] . \label{eq4-51}
\end{align}
Here, we neglected a term $\propto \delta(\Omega)$ by assuming $\Omega \neq 0$, and thus there is no linear term $({\mathcal E}_0/\bar{E})^1$ in the square brackets.  Also, we evaluated the $p_{\parallel}$-integration as $\int dp_{\parallel} = e\bar{E}T$.  This is because the momentum $p_{\parallel}$- and the time $x^0$-integration are related with each other in the presence of a constant electric field \cite{nik70}.  In fact, as we explained below Eq.~(\ref{eq3-42}), the particle production usually occurs at $x^0 = -p_{\parallel}/e\bar{E}$.  Thus, $e\bar{E} dx^0 = -dp_{\parallel}$ should hold, which yields $\int dp_{\parallel} = e\bar{E} \int dx^0 = e\bar{E}T$.

\begin{figure}[!t]
\includegraphics[clip, width=0.499\textwidth]{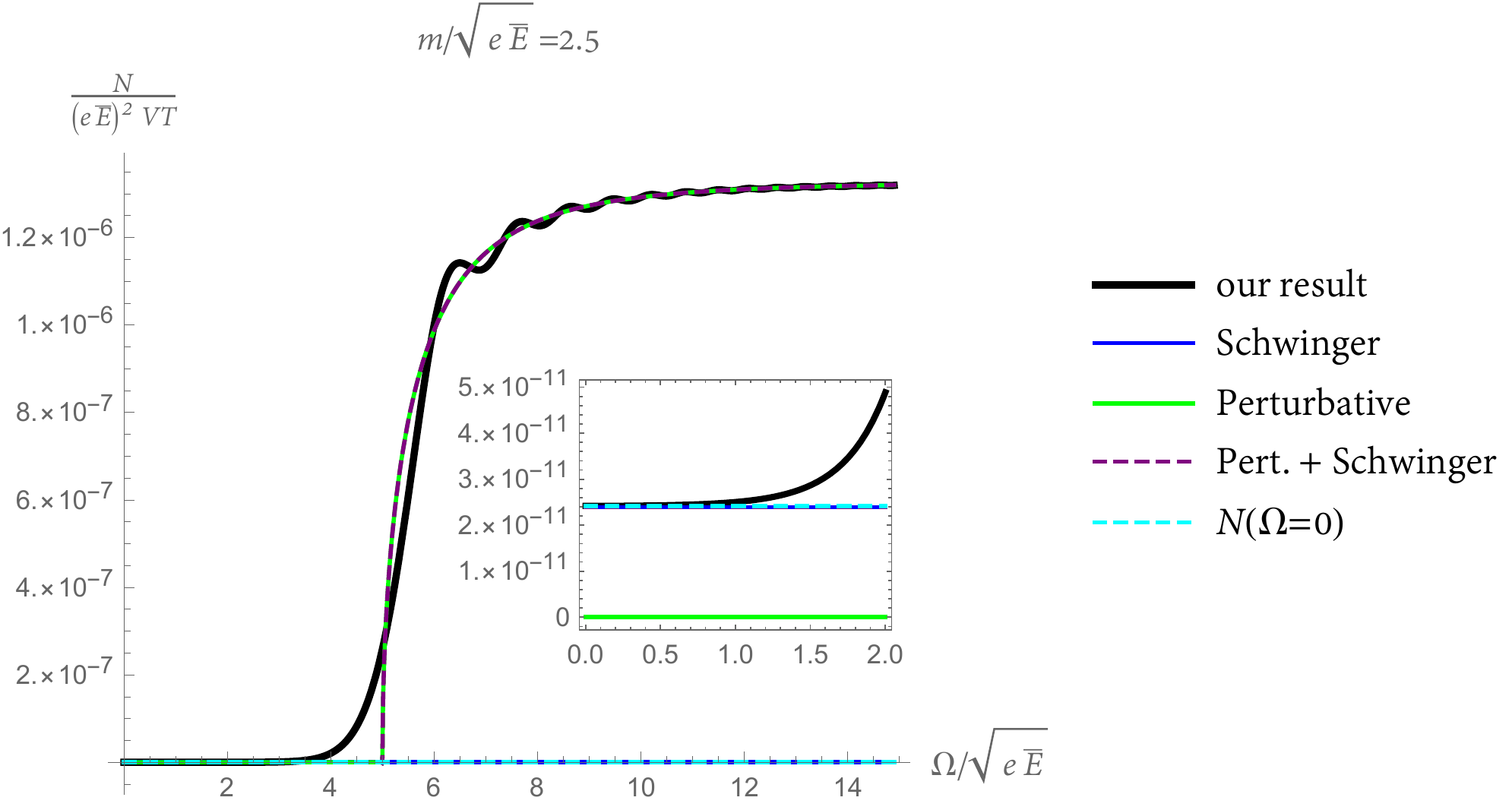}
\includegraphics[clip, width=0.499\textwidth]{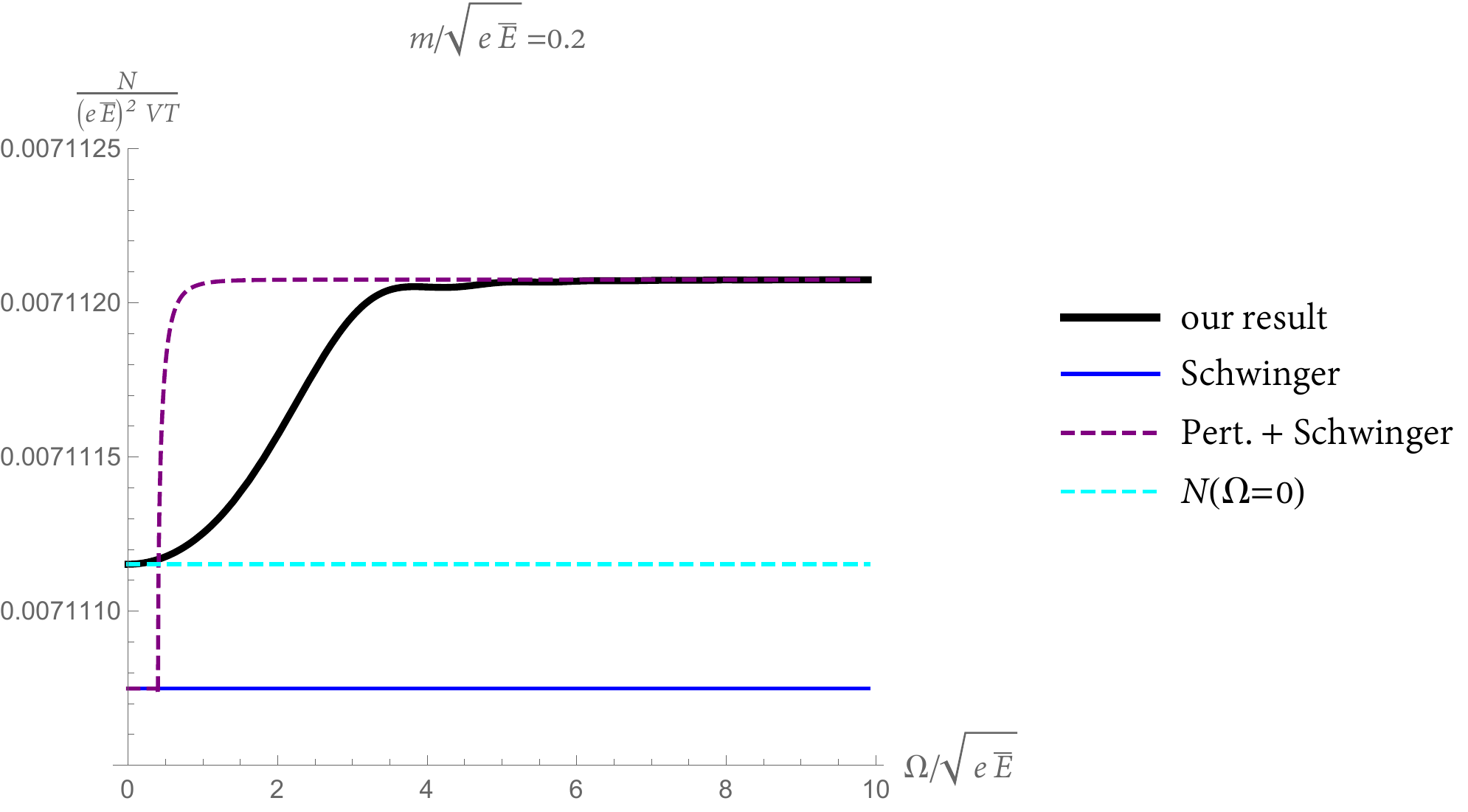}
\caption{\label{fig4} (color online) The frequency $\Omega$-dependence of the total number $N$ (thick black line) for sub-critical field strength $m/\sqrt{e\bar{E}} = 2.5$ (top) and super-critical one $m/\sqrt{e\bar{E}} = 0.2$ (bottom).   We fixed ${\mathcal E}_0$ as ${\mathcal E}_0 = \bar{E}/100$.  The green, blue, and dashed purple lines show the perturbative formula (\ref{eq4-53}), the Schwinger formula for the strong field (\ref{eq4-52}), and a superposition of them (\ref{eq4-54}).  The dashed cyan line shows the asymptotic value of $N$ at $\Omega = 0$ (Eq.~(\ref{eq4-55})).  }
\end{figure}

\begin{figure}[!t]
\includegraphics[clip, width=0.499\textwidth]{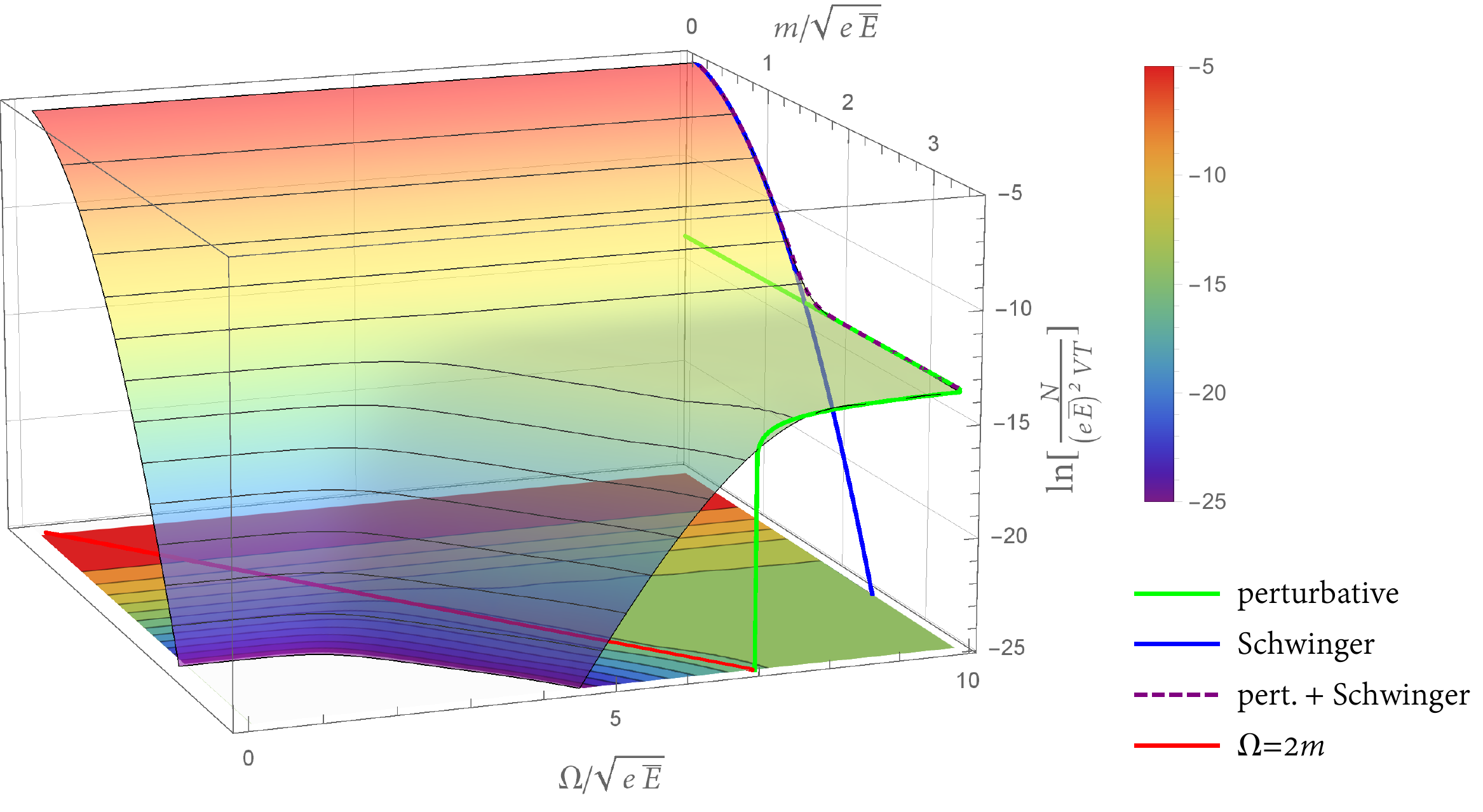}
\caption{\label{fig5} (color online) The total number $N$ as a function of the frequency $\Omega$ and mass $m$.  ${\mathcal E}_0$ is fixed ${\mathcal E}_0 = \bar{E}/100$.  The green, blue, and dashed purple lines show the perturbative formula (\ref{eq4-53}), the Schwinger formula for the strong field (\ref{eq4-52}), and a superposition of them (\ref{eq4-54}).  The red line at the bottom shows the threshold energy $\Omega = 2m$.  }
\end{figure}

The total number (\ref{eq4-51}) is plotted in Figs.~\ref{fig4} and \ref{fig5}.  For comparison, Schwinger's non-perturbative formula for the strong field $\bar{E}$ (the integration of Eq.~(\ref{eq3-43}))
\begin{align}
	N^{\rm (Sch)}	&= \sum_s \int d^3{\bm p}\; n^{\rm (Sch)}_{{\bm p},s} \nonumber\\
						&= (e\bar{E})^2 VT \times \frac{1}{4\pi^3} \exp\left[  -\pi \frac{m^2}{e\bar{E}}  \right]; \label{eq4-52}
\end{align}
the perturbative formula (the integration of Eq.~(\ref{eq3-46}))
\begin{align}
	N^{\rm (pert)}	&= \sum_s \int d^3{\bm p}\; n^{\rm (pert)}_{{\bm p},s} \nonumber\\
					& = (e\bar{E})^2 VT \times \frac{1}{48 \pi} \sqrt{1 - \frac{4m^2}{\Omega^2}} \nonumber\\
					&\quad \times \left( 2 + \frac{4m^2}{\Omega^2}  \right) \left| \frac{{\mathcal E}_0}{\bar{E}} \right|^2  \theta(\Omega - 2m); \label{eq4-53}
\end{align}
and a superposition of them (the integration of Eq.~(\ref{eq3-45}))
\begin{align}
	N^{\rm (Sch+pert)} = N^{\rm (Sch)} + N^{\rm (pert)}, \label{eq4-54}
\end{align}
are also plotted in the figures.  In Eq.~(\ref{eq4-54}), we dropped a cross term between Schwinger's non-perturbative production and the perturbative one because it is negligible in the limit of $T \to \infty$.  Notice that the perturbative formula (\ref{eq4-53}) has a cutoff at $\Omega = 2m$, which is the minimum energy to create a pair of particles from the vacuum $2\omega_{{\bm p}={\bm 0}} = 2m$.

One can roughly understand the parameter dependence of the total number $N$ (see Figs.~\ref{fig4} and \ref{fig5}) in terms of the interplay between the non-perturbative and perturbative particle production: For large frequency $\Omega/\sqrt{e\bar{E}} \gg 1$, the perturbative particle production by the weak field ${\mathcal E}$ occurs.  As the non-perturbative production by the strong field $\bar{E}$ is negligible for sub-critical field strength $e\bar{E} \lesssim m^2$, the perturbative process dominates the particle production and the total number is basically in agreement with the perturbative formula (\ref{eq4-53}).  Note, however, that there certainly exist small disagreements between them, which are nothing but the Franz-Keldysh effect and are discussed later in detail.  On the other hand, the non-perturbative production for super-critical field strength $e\bar{E} \gtrsim m^2$ becomes so abundant that the perturbative production just gives a small correction to the non-perturbative one.  Because of this small correction, the total number slightly deviates from the Schwinger formula for the strong field alone (\ref{eq4-52}) and it is consistent with the sum of the Schwinger formula and the perturbative formula (\ref{eq4-54}).  For small frequency $\Omega/\sqrt{e\bar{E}} \ll 1$, the perturbative particle production does not take place and the production becomes purely non-perturbative.  The total number becomes slightly larger than the Schwinger formula for the strong field alone (\ref{eq4-52}) because not only the strong field but also the weak field contributes to the non-perturbative particle production.  Indeed, Eq.~(\ref{eq4-51}) gives
\begin{align}
	N \xrightarrow{|\Omega| \to 0}{} &(e\bar{E})^2 VT \times \frac{1}{4\pi^3} \exp\left[  -\pi \frac{m^2}{e\bar{E}}  \right] \nonumber\\
	  & \times \left\{ 1 +   \left(  \frac{1}{2} + \frac{\pi}{2} \frac{m^2}{e\bar{E}} + \frac{\pi^2}{4} \left(\frac{m^2}{e\bar{E}}\right)^2 \right)   \left( \frac{{\mathcal E}_0}{\bar{E}} \right)^2 \right\}, \label{eq4-55}
\end{align}
which is actually larger than the Schwinger formula (\ref{eq4-52}) by the factor of ${\mathcal O} ( ( {\mathcal E}/\bar{E} )^2 )$.  Note that ${\mathcal O} ( ( {\mathcal E}/\bar{E} )^1 )$-correction is absent in Eq.~(\ref{eq4-55}), although corrections to the momentum distribution $n_{{\bm p},s}$ start from ${\mathcal O} ( ( {\mathcal E}/\bar{E} )^1 )$ (see Eq.~(\ref{eq3-42})).  This is because, for our monochromatic wave configuration, ${\mathcal O} ( ( {\mathcal E}/\bar{E} )^1 )$-correction in the total number becomes proportional to $\delta(\Omega)$ after $p_{\parallel}$-integration, and hence can be discarded for $\Omega \neq 0$.  For general field configurations, ${\mathcal O} ( ( {\mathcal E}/\bar{E} )^1 )$-correction in the total number cannot be a delta function and has a finite value even for $\Omega \neq 0$, so that the correction should start from ${\mathcal O} ( ( {\mathcal E}/\bar{E} )^1 )$.

\begin{figure}[!t]
\includegraphics[clip, width=0.45\textwidth]{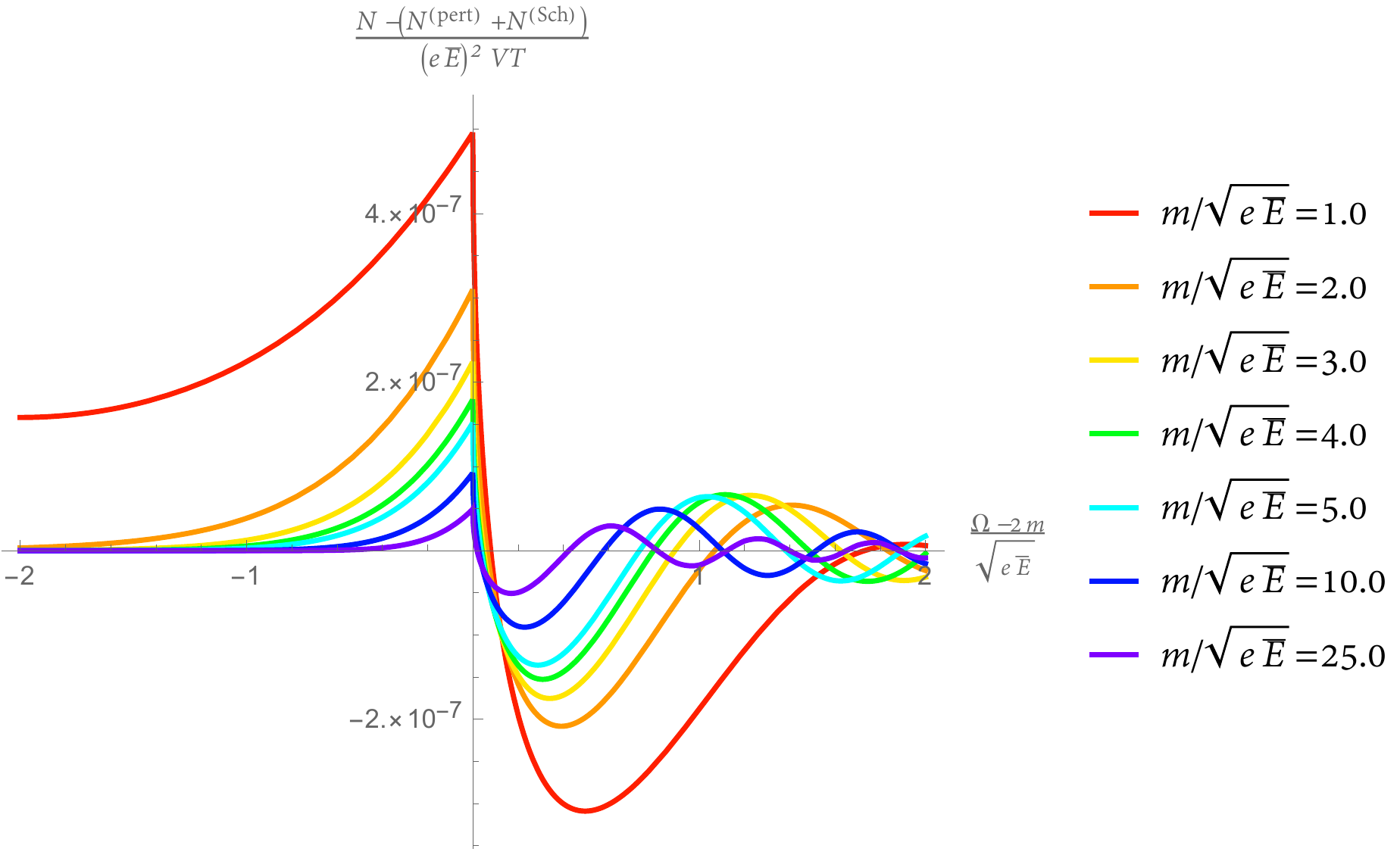}
\caption{\label{fig6} (color online) Difference between the total number and the sum of the Schwinger and the perturbative formula (\ref{eq4-54}), $N - (N^{\rm (Sch)} + N^{\rm (pert)})$, near the threshold $\Omega \sim 2m$.  ${\mathcal E}_0$ is fixed ${\mathcal E}_0 = \bar{E}/100$.  Different colors distinguish the strength of the strong field $\bar{E}$.  }
\end{figure}

As pointed out, although our result is basically in agreement with the perturbative formula (\ref{eq4-53}) or the sum with the Schwinger formula (\ref{eq4-54}) in the high frequency regime $\Omega/\sqrt{e\bar{E}} \gtrsim 1$, there certainly exist small disagreements between them.  The disagreements may be more clearly illustrated in Fig.~\ref{fig6}, in which absolute difference between our result and the sum of the Schwinger and the perturbative formula (\ref{eq4-54}), $\Delta N = N - (N^{\rm (Sch)} + N^{\rm (pert)})$, near the threshold $\Omega \sim 2m$ is plotted.  The disagreements are nothing but the Franz-Keldysh effect.  Namely, (i) the perturbative particle production occurs even below the threshold $\Omega \lesssim 2m$; (ii) the production number is slightly suppressed just above the threshold $\Omega \gtrsim 2m$; and (iii) the production number oscillates around the naive perturbative formula (\ref{eq4-53}) above the threshold $\Omega > 2m$.  The physical origin of this effect is completely the same as what we explained in Sec.~\ref{sec4-b1}, i.e., the change of wave function due to the quantum tunneling and/or reflection by the tilted gap in the presence of strong electric field.  Note that the difference $\Delta N $ corresponds to the change of the photon-absorption rate by the Franz-Keldysh effect in condensed matter physics, and is extensively measured in experiments \cite{boe59, wil60, wil62, vav60, mos61}.

An important point of the Franz-Keldysh effect is that it is suppressed only weakly by powers of $m/\sqrt{e\bar{E}}$ as can be seen from the figures.  This implies that, for a fixed sub-critical strong field, the Franz-Keldysh effect would be more manifest than the purely non-perturbative particle production mechanism (the Schwinger mechanism), which is strongly suppressed by an exponential of $(m/\sqrt{e\bar{E}})^{-1}$.  As an example, let us consider a sub-critical strong electric field with $\sqrt{e\bar{E}} \lesssim \sqrt{10^{-2} \times eE_{\rm cr}} = 10^{-1} \times m_{\rm e} = {\mathcal O}(10\;{\rm keV})$, $V = {\mathcal O}((1\;\mu{\rm m})^3)$, and $T = {\mathcal O}( 1\;{\rm fsec}$), which are typical values within the current laser technologies \cite{pia12}.  Although it may still be difficult within the current laser technologies to realize coherent electric fields with high-frequency exceeding the electron mass scale, let us assume for now that we have such a high frequency electric field with weak strength.  Then, $N^{\rm (Sch)}$ can be estimated as $N^{\rm (Sch)} \sim (4 \times 10^{-118}) \times T[{\rm fsec}]$, which is, obviously, an extremely small number.  On the other hand, the Franz-Keldysh effect predicts that there should be a difference in the production number $\Delta N = N - (N^{\rm (pert)} + N^{\rm (Sch)}) \sim N - N^{\rm (pert)}$ with and without the strong electric field.  From Fig.~\ref{fig6}, $\Delta N$ can be estimated as $\Delta N \sim (1 \times 10^{14}) \times T[{\rm fsec}]$ (where $N^{\rm (pert)} \sim (1 \times 10^{14\sim15}) \times T[{\rm fsec}]$) for $ {\mathcal E}_0 = 10^{-2} \times \bar{E} = 10^{-4} \times E_{\rm cr}$.  This is a huge number.  Note that this order estimate may be valid when $T$ is not so long $T\lesssim 10^{-4}\;{\rm fsec}$, for which back-reaction may safely be neglected\footnote{In order to make a more realistic estimate, it is important, in particular for longer $T$, to consider back-reaction effects of the particle production to the electric field.  In general, back-reaction becomes important when the energy of produced particles becomes comparable to that of the electric field \cite{mat83, bul04, tan09}.  Back-reaction would deplete the electric field, so that one has to inject additional energy (or sources) to the field in order to maintain the field strength; otherwise the present order estimate for $\Delta N$ fails to reproduce the actual production number.  In the present parameter choice, the energy of produced particles ${\mathsf E}_{\rm e} \sim m_{\rm e} N \sim (1 \times 10^{14 \sim 15}) \times m_{\rm e} \times T[{\rm fsec}]$ becomes comparable to the field's energy ${\mathsf E}_{\rm field} \sim V\bar{E}^2/2 \sim (1 \times 10^{15}) \times m_{\rm e}$ if $T \sim 1\;{\rm fsec}$.  Also, the perturbative estimate (\ref{eq4-53}) may be valid for $m_{\rm e} N^{({\rm pert})} \lesssim {\mathcal E}_0^2/2\ \Rightarrow\ T \lesssim (1\times 10^{+2}) \times m_{\rm e}^{-1} \sim 1\times10^{-4}\;{\rm fsec}$.  

Note that one can apply electric fields with short duration $\Delta T$ for many times $n$.  Back-reaction for each production process can safely be neglected as long as $\Delta T$ is sufficiently short, and hence the total production number should be consistent with the present estimate without back-reaction with $T=n\Delta T$.  

We leave the back-reaction problem as a future work.  }.  Therefore, the Franz-Keldysh effect is actually more manifest than the Schwinger mechanism, and would be testable by experiments if we have a weak electric field with high-frequency even if the strong field is not so strong.  Note that the Franz-Keldysh effect depends on ${\mathcal E}$ quadratically.  Thus, the Franz-Keldysh effect may still be manifest even for very weak ${\mathcal E}$ (e.g. ${\mathcal E}_0 \sim 10^{-9} \times {\bar E} = 10^{-11} \times E_{\rm cr} $ still gives a significant difference $\Delta N \sim 1\times T[{\rm fsec}]$).

\section{Summary and Discussion} \label{sec4}

We studied an analog of the Franz-Keldysh effect in QED, and the interplay between the non-perturbative (the Schwinger mechanism) and the perturbative particle production in the presence of a strong slow field and a weak perturbation on top of it. 

In Sec.~\ref{sec2}, we derived a general formula for the produced number of particles.  Firstly, we used the retarded Green function technique to solve the Dirac equation perturbatively with respect to the weak perturbation, while the interactions due to the strong field are treated non-perturbatively.  We, then, employed the canonical quantization procedure in the presence of the strong field, and directly computed the in-vacuum expectation value of the number operator.  The obtained formula (\ref{eq2-28}) is written in terms of bi-linears between positive/negative frequency mode functions at in- and out-states, which are fully dressed by the strong field.  This dressing enables us to study the Franz-Keldysh effect, which is a cooperative effect between the non-perturbative particle production mechanism due to the strong field and the perturbative one due to the weak perturbation.  Also, the formula (\ref{eq2-28}) is valid no matter how fast or slow the weak perturbation is as long as the perturbation is sufficiently weaker than the strong field.  Thus, the formula (\ref{eq2-28}) is able to describe the interplay between the perturbative particle production and the non-perturbative one (the Schwinger mechanism) with changing characteristic time-scale of the perturbation.

In Sec.~\ref{sec3}, we considered a specific field configuration to discuss features of the particle production in more detail.  To be concrete, we assumed that the strong field and the weak perturbation are given by a constant homogeneous electric field and a monochromatic wave, respectively.  In this configuration, we analytically evaluated Eq.~(\ref{eq2-28}) without any approximations, and explicitly demonstrated how the interplay and the Franz-Keldysh effect occur.  In particular, we found that the Franz-Keldysh effect significantly affects the perturbative particle production mechanism: (i) the perturbative particle production occurs even below the threshold energy; (ii) the perturbative production becomes the most efficient just above the threshold energy; and (iii) a characteristic oscillating pattern appears in the production number above the threshold energy.  We argued that these changes are naturally explained in terms of changes of wave function of electrons in the Dirac sea due to quantum tunneling/reflection in the presence of the strong electric field.  In addition, we claimed that (i) is essentially the same as the dynamically assisted Schwinger mechanism, and, therefore, one may understand the Franz-Keldysh effect in QED as a generalization of the dynamically assisted Schwinger mechanism.

We also found that the Franz-Keldysh effect is suppressed only weakly by powers of the critical field strength of QED.  In other words, for a fixed sub-critical strong field, the Franz-Keldysh effect would be more manifest than the purely non-perturbative particle production mechanism (the Schwinger mechanism), which is strongly suppressed by an exponential of the critical field strength.  Unfortunately, however, it is still difficult within the current laser technologies to realize coherent electric fields with high-frequency exceeding the electron mass scale.  Nevertheless, once such high-frequency electric fields are realized in future laser experiments and/or other physical systems with strong fields such as ultra-relativistic heavy ion collisions, the Franz-Keldysh effect would take place, which in turn leaves significant experimental signatures.  One of the possible experiments is to measure the difference $\Delta N$ between the number of produced particles from the vacuum by a weak monochromatic wave with and without a strong electric field as suggested in Fig.~\ref{fig6}.  This is an analog of ``modulation spectroscopy,'' which is actually used in the area of condensed matter physics to detect the Franz-Keldysh effect in semi-conducting materials \cite{car69}.  We expect that $\Delta N$ significantly deviates from zero near the threshold and exhibits characteristic patterns in the frequency dependence, e.g., an exponential tail below the threshold; a very sharp peak at the threshold; and an oscillation above the threshold.

The Franz-Keldysh effect would serve as a powerful tool to study non-perturbative aspects of QED.  In particular, it would be very useful to investigate the vacuum structure of QED.  This is because the Franz-Keldysh effect occurs due to changes of wave function of particles filling the Dirac sea as we explained in Sec.~\ref{sec4-b1}, and the changes are directly related to actual observables, i.e., frequency-dependence of the produced particle number.  In fact, in the area of condensed matter physics, the Franz-Keldysh effect is experimentally used to precisely determine band structure of semi-conducting materials, and is very successful \cite{car69}.

There are several possible future directions for this work.  One direction is to consider more realistic field configurations, e.g., spatially inhomogeneous fields and/or perturbations; polarized perturbations; and inclusion of strong magnetic fields.  Although we concentrated on the simplest situation in Sec.~\ref{sec3} (i.e., a constant homogeneous strong electric field and a weak monochromatic wave) for simplicity, our general perturbative formalism developed in Sec.~\ref{sec2} can be directly applicable to these more general situations.  Note that a similar perturbative approach was used to study these more realistic situations in the context of the dynamically assisted Schwinger mechanism \cite{tor17, tor18}.  This extension is not only important for actual experiments, but also interesting from a phenomenological point of view.  For example, recently it is argued in the context of the dynamically assisted Schwinger mechanism that spatial inhomogeneous perturbations dramatically change the production number \cite{sch16, cop16}.  As the dynamically assisted Schwinger mechanism can be understood as a part of the Franz-Keldysh effect, we expect that similar changes should appear in the Franz-Keldysh effect as well.  Inclusion of strong magnetic fields should also have significant impacts on the Franz-Keldysh effect because the Landau quantization comes into play.  In fact, non-trivial changes are discussed in the area of condensed matter physics as a result of the interplay between strong electric and magnetic fields \cite{aro63, aro66, zaw66, zaw73}.

Another direction we would like to mention is applications to phenomenology.  In particular, an application to heavy ion physics may be interesting.  Just after a collision of ultra-relativistic heavy ions at RHIC and/or LHC, there appears very strong chromo-electromagnetic field (sometimes called ``glamsa''), whose typical strength is ${\mathcal O}(1\;{\rm GeV})$ \cite{low75, nus75, kov95a, kov95b, lap06}.  In addition to the glasma, there also exist jets, which are made up of high-energetic partons originating from initial hard collisions.  Although the typical energy scale of jets are ${\mathcal O}(100\;{\rm GeV})$, there are thousands of low-energetic jets ($\sim$ a few GeV), which are called mini-jets.  Thus, one may regard the system just after a collision as a superposition of strong field (glamsa) and weak perturbations on top of it ((mini-)jets).  This is essentially the same situation that we discussed in this paper.  Thus, the Franz-Keldysh effect may take place.  The Franz-Keldysh effect may change parton splitting functions, which might soften jet spectrum and help mini-jets to thermalize.

\section*{Acknowledgements}

The author would like to thank K.~Hattori, X.-G.~Huang, and the members of RIKEN iTHEMS STAMP working group for fruitful discussions.   The author is supported by NSFC under Grant No.~11847206.

\end{document}